\documentclass[11pt,a4paper]{article} 
\pdfoutput=1
\usepackage{jheppub}

\usepackage{amsmath, amssymb}
\usepackage{mathpazo}
\usepackage{mathrsfs}
\usepackage{array,arydshln}

\usepackage{graphicx,epsfig}
\usepackage{epic}
\usepackage{color}
\usepackage{youngtab}
\usepackage{float}

\newcommand{\be}{\begin{equation}}
\newcommand{\ee}{\end{equation}}
\newcommand{\ba}{\begin{eqnarray}}
\newcommand{\ea}{\end{eqnarray}}

\newcommand{\mc}{\mathcal }

\newcommand{\N}{\mathcal{N}}

\newcommand{\mk}{\mathfrak}
\newcommand{\hyper}[3]{{}_{3}F_{2}\left(\left.\begin{array}{c} #1 \\ #2 \end{array}\right | #3\right)}
\newcommand{\hs}{\mbox{hs}[\lambda]}

\newcommand{\figref}[1]{Fig.~(\ref{fig:#1})}

\def\XXint#1#2#3{{\setbox0=\hbox{$#1{#2#3}{\int}$}
     \vcenter{\hbox{$#2#3$}}\kern-.5\wd0}}

    \newcommand{\beq}{\begin{equation}}
    \newcommand{\eeq}{\end{equation}}
    \newcommand\beqa{\begin{eqnarray}}
    \newcommand\eeqa{\end{eqnarray}}

\title{Analysis of higher spin black holes with spin-4 chemical potential}

\author[a]{Matteo Beccaria} 
\author[a]{, Guido Macorini}

\affiliation[a]{Dipartimento di Matematica e Fisica Ennio De Giorgi,\\
Universit\`a del Salento \& INFN, Via Arnesano, 73100 Lecce, 
Italy} 
                     
\emailAdd{matteo.beccaria@le.infn.it}
\emailAdd{guido.macorini@le.infn.it}

\abstract{
We consider the $AdS_{3}/CFT_{2}$ duality between certain coset WZW theories at large central charge
and Vasiliev 3D higher spin gravity with a single complex field. On the gravity side, we discuss a higher spin
black hole solution with chemical potential  coupled to the spin-4 charge. We compute the perturbative expansion of the higher spin charges and of the partition function at high order in the chemical potential. The result is obtained
with its exact dependence on the parameter $\lambda$ characterising the symmetry algebra $\hs$. The cases of $\lambda=0,1$ are successfully compared with a CFT calculation. The special point $\lambda=\infty$, 
the Bergshoeff-Blencowe-Stelle limit,  is also solved in terms of the exact generating function for the partition function.
The thermodynamics of both the spin-4 and the 
usual spin-3 black holes is studied in order to discuss the $\lambda$ dependence of the BTZ critical temperature $T_{\rm BTZ}(\lambda)$.
In the spin-3 case, it is shown that $T_{\rm BTZ}(\lambda)$ converges for large $\lambda$ to the critical point of the 
$\lambda=\infty$ known partition function previously found by the authors. 
In the spin-4 black hole,  the picture is qualitatively similar and  $T_{\rm BTZ}(\infty)$ is accurately determined by various numerical methods.
}

\allowdisplaybreaks

\begin{document} \maketitle

\bigskip

\section{Introduction}

Gravity theories with interacting higher spin gauge fields plays an important role in the study of AdS/CFT 
correspondence \cite{Maldacena:1997re,Witten:1998qj}. With major insight, the authors of  \cite{Gaberdiel:2010pz}
proposed Vasiliev higher spin theory on $AdS_{3}$ \cite{Vasiliev:1995dn,Vasiliev:1996hn}  to be 
dual to a class of coset  WZW conformal theories in certain large-$N$ limit. The gravity side is a topological 
three dimensional theory with infinite dimensional gauge symmetry $\mk{hs}[\lambda]\oplus\mk{hs}[\lambda]$. 
As a consequence of the duality, the conformal side has chiral algebra $\mc W_{\infty}[\lambda]$. 

\vskip 5pt
The gravity theory in the bulk is known to admit  black hole solutions 
with non zero  values of the higher spin charges \cite{Gutperle:2011kf,Kraus:2011ds,Ammon:2012wc}.  
These black holes are a tool to understand and probe AdS/CFT duality. Besides, they are interesting in their own because the invariance under diffeomorphism is enlarged to a higher spin symmetry. Remarkably, it is possible to describe the 
thermodynamics of the black hole solutions in quite explicit terms \cite{Gutperle:2011kf,Ammon:2011nk}.
In particular, the entropy can be computed from the partition function. Non trivial values of the higher spin charges
are induced by introducing chemical potentials coupled to the  higher spin currents. In the AdS/CFT correspondence, 
such black holes are   interpreted as states of a dual CFT deformed by an irrelevant operator determined by which 
chemical potentials are turned on. The counting of microscopic states can be performed in the CFT by exploiting
the chiral $\mc W$-symmetry of the conformal theory. The simplest theoretical framework is the spin-3 black hole
where there is a chemical potential sourcing the spin-3 charge. For this problem, the 
$\mc O(\alpha^{8})$ expansion of the partition function in powers of the spin-3 charge 
chemical potential $\alpha$ \cite{Kraus:2011ds} has been confirmed by the computations in 
\cite{Gaberdiel:2011zw,Gaberdiel:2012yb,Gaberdiel:2013jca} at generic values of the parameter $\lambda$.
We remind that $\lambda$ has the physical role of parametrising  a curve of inequivalent AdS vacua of the theory.

\vskip 5pt
Actually, the dependence on $\lambda$ is quite interesting and three special values are particularly intriguing.
At $\lambda=0,1$  the dual conformal theory is free and the partition function can be computed 
at all orders. At  $\lambda=\infty$, the Bergshoeff-Blencowe-Stelle limit \cite{Bergshoeff:1989ns}, many simplifications occur and the limit is non-trivial. The partition function in this
regime has been computed in closed form in \cite{Beccaria:2013dua} and displays a critical point whose physical meaning will be elucidated in this paper. Also, special resummation properties have been 
discovered in the study of the scalar correlator in the spin-3 background \cite{Beccaria:2013yca}.

\vskip 5pt
In this paper, we introduce another simple theoretical model that can be studied with the same techniques 
developed for the spin-3 black hole. It is a black hole with a chemical potential for the spin-4 charge. 
Such a solution has been considered in fixed low $\N$ gravity in  \cite{Tan:2011tj,Chen:2012pc,Chen:2012ba,Ferlaino:2013vga}. Here, we shall present 
a detailed study of such a solution in $\hs$ by uplifting the solution in $\mk{sl}(\N)$ gravity at generic $\N$
and identifying $\N$ with $\lambda$. In particular, we shall present a high order calculation of the perturbative
expansion of the higher spin charges and the partition function. For the spin-4 black hole, we shall prove 
agreement with CFT at $\lambda=0,1$ thus providing a novel partial check of $AdS_{3}/CFT_{2}$ duality.
Also, we shall  discuss the properties of the $\lambda=\infty$ point finding, in particular, the exact generating function
of the partition function.

\vskip 5pt
As a matter of fact, the truncation of the $\hs$ theory to $\mk{sl}(\N)$ gravity at $\lambda=\N$ is an important tool in the study of the dependence on $\lambda$ which is typically smooth and can be inferred from the analysis of the case of integer $\lambda$.
In particular,  the thermodynamics of higher spin black 
holes in $\mk{sl}(\N)$ gravity has been studied in various papers (see for instance \cite{David:2012iu}) with much
effort in the $\N=3$ case with spin-3 chemical potential. The perturbative BTZ branch of the solution admits a critical temperature where a first order transition is expected to occur and the BTZ branch ceases to exist. In the past, the dependence of this temperature on $\N$ has not been studied in details. Here, we  perform such a study in both the spin-3 and spin-4 black holes.
We shall prove that the critical BTZ temperature exists for all $\N$ and converges as $\N\to\infty$. In the spin-3 black hole, we  provide accurate numerical results to show that its limit value can be identified with the critical point of the large $\lambda$  partition function. In the spin-4 black hole, we  demonstrate that the pattern is the same and that the critical BTZ temperatures
converges to  a critical point of the large $\lambda$ partition function related to the finite radius of convergence of 
its perturbative expansion.

The plan of the paper is the following. 
In Sec.~(\ref{sec:intro}), we review the construction of higher spin black holes in $\mk{sl}(\N)$ Chern-Simons gravity.
In Sec.~(\ref{sec:spin4}), we present the spin-4 black hole solution in $\mk{sl}(\N)$ with generic $\N$, thus essentially in $\hs$.
In Sec.~(\ref{sec:CFTtest}), we show the agreement with the gravity partition function and the CFT calculation at $\N=0,1$.
In Sec.~(\ref{sec:CriticalTemperature}), we present the analysis of the BTZ critical temperature in both the spin-3
and spin-4 black holes with emphasis on its $\N\to\infty$ limit. Various appendices collect technical data and complementary discussions.

\section{Higher spin black holes in $\mk{sl}(\mc N)\oplus\mk{sl}(\mc N)$ Chern-Simons gravity}
\label{sec:intro}

Einstein gravity with a negative cosmological constant can be recast in the form of 
a $SL(2)\times SL(2)$ Chern-Simons theory \cite{Achucarro:1987vz,Witten:1988hc}. 
The action is a functional of the $\mathfrak{sl}(2)$-valued 1-forms $A$ and $\overline A$ 
\be
S = S_{\rm CS}(A)-S_{\rm CS}(\overline A),
\ee
with
\be
S_{\rm CS}(A) = \frac{k}{4\,\pi}\int\mbox{Tr}\bigg(A\wedge dA+\frac{2}{3}\,A\wedge A\wedge A\bigg).
\ee
The Chern-Simons level $k$,  the Newton constant $G_{\rm N}$, and the $AdS_{3}$ radius $\ell_{\rm AdS}$ 
are related by  $k = \ell_{\rm AdS}/(4G_{\rm N})$.
The extension from $\mathfrak{sl}(2)$ to $\mk{sl}(\N)$, with integer $\N\ge 3$, is particularly interesting. It describes a gravity theory where the graviton
is supplemented by a tower of symmetric tensor fields with spin $s = 3,4,...,\N$. 
Remarkably, this theory can be viewed as a truncation of the Chern-Simons action based on the 
infinite dimensional higher spin algebra $\mk{hs}[\lambda]$ where $\lambda$ is a positive real parameter 
determining the gravitational couplings among the higher spin fields \cite{Prokushkin:1998bq,Campoleoni:2010zq}.
Upon the choice $\lambda=\mc N$, we recover the $\mk{sl}(\N)$ theory. 

\bigskip
Higher spin black holes can be constructed as a suitable generalisation of the BTZ black hole found in \cite{Banados:1992wn,Banados:1992gq} in $\mk{sl}(3)$ gravity. The (holomorphic) connection of the BTZ black-hole reads 
\be
A = \bigg(e^{\rho}V_{1}^{2}-\frac{2\pi}{k}\,\mc L\,e^{-\rho}\,V_{-1}^{2}\bigg)\,dx^{+}+V_{0}^{2}\,d\rho, 
\ee
where $\rho, x^{\pm}\equiv t\pm \varphi$ are the space-time coordinates and the conserved charge $\mc L$
is a linear combination of the conserved mass and angular momentum charges. The operators $V^{2}_{s}$
are generators of $\mk{sl}(2)$ principally embedded in $\mk{sl}(3)$ (see App.~(\ref{app:notation}) for the notation).
It is convenient to introduce complex coordinates on the fixed-$\rho$ slices with the identification $z\sim z+2\pi \tau$. The  modular parameter $\tau$ 
depends on the temperature and angular velocity of the black hole \cite{Ammon:2012wc}. The holonomy
of the gauge connection around the above Euclidean time circle is (notice that here $A_{-}=0$)
\be
\omega = 2\,\pi\,(\tau A_{+}-\overline \tau A_{-}).
\ee
The holonomy  has to be trivial if we want to smoothly close the circle at the horizon. This smoothness condition leads to
the spin-2 charge 
\be
\mc L = -\frac{k}{8\,\pi\,\tau^{2}},
\ee
and can be alternatively stated in a gauge invariant way as a constraint on the holonomy eigenvalues.

Similar solutions can be constructed in the  $\mk{sl}(\N)$ theory 
according to the recipe described in \cite{Kraus:2011ds}.
As a first step, one performs a suitable gauge transformation that removes from the connections any $\rho$ dependence
and also sets to zero the $\rho$ component \cite{Coussaert:1995zp}
\be
A = b^{-1}a b + b^{-1}db.
\ee
The $'+'$ component of the transformed connection $a$ is then written in the form 
\be
a_{+} = V^{2}_{1}-\frac{2\,\pi\,\mc L}{k}\,V^{2}_{-1}+\sum_{s=3}^{\infty}\nu_{s}\,\mc W_{s}\,V^{s}_{-s+1},
\ee
where $V^{s}_{m}$ are $\mk{sl}(\N)$ generators, $\nu_{s}$ are normalisation constants, and $\mc W_{s}$ are higher spin (constant) charges. The flatness condition $[a_{+},a_{-}]=0$ is trivially solved by postulating the following expression for
the component $a_{-}$
\be
a_{-} = \sum_{s=1}^{\infty} \mu_{s}\,\left. a_{+}^{s}\right|_{\rm traceless}+{\rm subleading}.
\ee
The physical meaning of the constants $\mu_{s}$ is that of chemical potentials acting as  sources for the higher spin charges appearing in $a_{+}$ (the subleading terms will be discussed later). This interpretation can be supported by a more constructive approach where one shows that, for non constant charges $\mc W_{s}=\mc W_{s}(z, \overline z)$ and $\mu_{s}=\mu_{s}(z, \overline z)$,  the bulk equations of motion reduce to the Ward identities of the asymptotic chiral algebra in presence of 
deformations associated with the higher spin fields \cite{Gutperle:2011kf}. The standard way to fix the charges as functions of the chemical potentials is to impose smoothness of the horizon in the form of the following (infinite) set of 
equations \cite{Gutperle:2011kf,Ammon:2011nk}
\be
\mbox{Tr}(\omega^{n}) = \mbox{Tr}(\omega_{\rm BTZ}^{n}), \qquad\quad 
\omega_{\rm BTZ} = 2\,\pi\,\tau\,\bigg(V^{2}_{1}+\frac{1}{4\,\tau^{2}}\,V^{2}_{-1}\bigg).
\ee
These holonomy conditions must be consistent with thermodynamics. In other words, we want to interpret the
black hole solution as a saddle point contribution to a microscopic partition function of the form 
(holomorphic part only)
\be
Z(\tau, \mu_{2}, \dots) = \mbox{Tr}\bigg[e^{4\pi^{2}i(\tau \widehat{\mc L}
+\sum_{s}\mu_{s} \widehat{\mc W_{s}})}\bigg] = e^{S+4\pi^{2}\,i\,(\tau {\mc L}
+\sum_{s}\mu_{s} {\mc W_{s}})},
\ee
where $S$ is the entropy in the holomorphic formalism~\footnote{
It has been shown in \cite{Perez:2013xi,deBoer:2013gz} that the associated entropy does not 
agree with the canonical entropy. The discrepancy is clarified \cite{Compere:2013gja,Compere:2013nba}
 where 
thermodynamical variables are defined  so that the canonical partition function has a natural CFT interpretation. }. By
consistency, the following integrability relations must hold
\be
\label{eq:integrability}
\frac{\partial\mc L}{\partial\mu_{s}} = \frac{\partial\mc W_{s}}{\partial\tau}.
\ee
Actually, these conditions are quite strong. We shall exploit them in the analysis of the spin-4 black hole studied in this paper. In particular, they will fix the structure of the various terms in the connection including the 
subleading terms.

\subsection{The spin-3 black hole}

The previous discussion can be made definite by considering the spin-3 black hole introduced in  \cite{Kraus:2011ds}.
This is a solution with a non trivial spin-3 charge $\mc W$ associated with the chemical potential $\alpha$. The solution has been discussed for the $\hs$ invariant theory, and can be truncated to a solution in $\mk{sl}(\N)$ by taking 
$\lambda=\N$. The detailed form of the connection
is 
\ba
a_{+} &=& V^{2}_{1}-\frac{2\,\pi\,\mc L}{k}\,V^{2}_{-1}-\kappa(\N)\,\frac{\pi\,\mc W}{2\,k}\,V^{3}_{-2}+
\mc J_{4}\,V^{4}_{-3}+\mc J_{5}\,V^{5}_{-4}+\cdots, \\
a_{-} &=& \mu\,\kappa(\N)\,\left. a_{+}^{2}\right|_{\rm traceless},\qquad \overline\tau\,\mu = \alpha.
\ea
where the normalisation $\kappa(\N) = \sqrt\frac{20}{\N^{2}-4}$ simplifies the comparison with the $\mk{sl}(3)$
results of  \cite{Gutperle:2011kf,Ammon:2011nk} and is also adopted in  \cite{Kraus:2011ds}. The analysis 
of  \cite{Kraus:2011ds,Beccaria:2013dua} determined the high order expansion of the charges as power series in $\alpha$ whose coefficients are explicit functions of $\N$. The integrability conditions (\ref{eq:integrability})
are satisfied and the partition function $\log Z$ can be computed.
Remarkably, it is possible to take the $\N\to\infty$ limit of various quantities in closed form. In particular, 
it was found that ($k=1$)
\be
\log Z_{\N=\infty}(\tau, \alpha) = \frac{3\,i\,\pi\,\tau^{3}}{160\,\alpha^{2}}\,\bigg[
\hyper{-\frac{3}{4}, \ -\frac{1}{2}, \ -\frac{1}{4}}{\frac{1}{3}, \ \frac{2}{3}}{-\frac{5120\,\alpha^{2}}{81\,\tau^{4}}}
-1
\bigg].
\ee
From this expression, the exact spin-2 and spin-3 charges are
\be
\mc L = \frac{1}{4\,\pi^{2}\,i}\,\frac{\partial\log Z}{\partial\tau},\qquad
\mc W = \frac{1}{4\,\pi^{2}\,i}\,\frac{\partial\log Z}{\partial\alpha}.
\ee
Also, we have 
\ba
\N^{2}\, \mc J_{4}(\alpha, \tau) &\stackrel{\N\to\infty}{=}& \frac{21}{3200\,\alpha^{4}}\,\bigg[
 -3\,\tau^{4}\,
\hyper{-\frac{3}{4}, \ -\frac{1}{2}, \ -\frac{1}{4}}{\frac{1}{3}, \ \frac{2}{3}}{-\frac{5120\,\alpha^{2}}{81\,\tau^{4}}}\nonumber \\
&&+40\,\alpha^{2}\,
 \hyper{\frac{1}{4}, \ \frac{1}{2}, \ \frac{3}{4}}{\frac{4}{3}, \ \frac{5}{3}}{-\frac{5120\,\alpha^{2}}{81\,\tau^{4}}}+40\alpha^{2}+3\,\tau^{4}
 \bigg].
 \ea
 These expressions show that there is a critical (branch point) value located at 
 \be
 \label{eq:spin3-alphac}
\left |\frac{\alpha}{\tau^{2}} \right| = \frac{9}{32\,\sqrt 5}.
 \ee
 In Sec.~(\ref{sec:CriticalTemperature}), we shall explore and clarify the physical interpretation of this singularity.

\section{The spin-4 black hole}
\label{sec:spin4}

A black-hole solution with a spin-4 source has been considered in 
\cite{Tan:2011tj,Chen:2012pc,Chen:2012ba,Ferlaino:2013vga}
at $\N=4,5$. Here, we propose the general form of the solution for generic $\N$, including the uplift to $\mbox{hs}[\lambda=\N]$.
The connection is 
\ba
a_{+} &=& V^{2}_{1}-\frac{2\,\pi\,\mc L}{k}\,V^{2}_{-1}+\frac{\pi}{3\,k}\,\frac{7}{\mc N^{2}-9}\,\mc J_{4}\,V^{4}_{-3}+\sum_{n=3}^{\infty}\mc J_{2n}\,V^{2n}_{-2n+1}, \\
a_{-} &=& \mu\,\bigg(\kappa(\mc N)^{2}\,\left. a_{+}^{3}\right|_{\rm traceless}-\frac{8\,\pi\,\mc L}{k}\,\frac{3\,\mc N^{2}-7}{\mc N^{2}-4}\,a_{+}\bigg), \qquad \overline \tau \,\mu = \alpha.
\ea
This expressions contains arbitrary normalisations of the charges and the chemical potential. We choose the coefficient
of the $a_{+}^{3}$ term in the $'-'$ component  to be $\kappa^{2}$ in analogy with the spin-3 case. This choice will guarantee in the end a smooth limit for $\N\to\infty$. Once these normalisations are fixed, everything is determined by the integrability conditions. They determine, in particular, the peculiar coefficient in front of the linear term $\sim a_{+}$ in $a_{-}$. As a final remark, we notice that all the odd spin charges are zero in this solution. This is due to the fact that the source is associated with the (even) spin-4 field. 

\subsection{Perturbative expansion}

We have solved the holonomy equations for our solution and the general form of the expansion of the higher spin charges 
in powers of $\alpha$ turns out to be ($\mc J_{2}\equiv \mc L$)
\be
\label{eq:expansions}
\mc J_{s} = \sum_{n=\frac{s}{2}-1}^{\infty}\mc J_{s,n}(\N)\,\frac{\alpha^{n}}{\tau^{3n+s}}, \qquad s=2, 4, 6, \dots.
\ee
The functions $\mc J_{s,n}(\N)$ are rational functions of degrees $[d_{1}:d_{2}]$ where
\be
\left[d_{1}:d_{2}\right] = \left\{\begin{array}{ll}
\left[2n-2:2n-2\right], & \quad s=2, \\
\left[2n:2n\right], & \quad s=4, \\
\left[2n-s+2:2n\right], & \quad s\ge 6.
\end{array}\right.
\ee
This means that $\mc L$ , $\mc J_{4}$, and $\N^{s-2}\,\mc J_{s}$ ($s\ge 6$) admit a smooth non trivial limit as
 $\N\to\infty$. Many terms $\mc J_{s,n}$ are reported in App.~(\ref{app:ratfuncs}). Also, 
 some features of the rational functions $\mc J_{s,n}$ are discussed in App.~(\ref{app:zeroes}).
 Here, we just write the first terms of the expansions of the spin 2, 4, 6 charges
 \ba
 \mc L &=& -\frac{k}{8 \pi  \tau ^2}-\frac{3 k
   \left(\mathcal{N}^2-9\right)}{2  \pi  \left(\mathcal{N}^2-4\right) }\,\frac{\alpha^{2}}{\tau^{8}}
   +\frac{10 k \left(\mathcal{N}^4-28 \mathcal{N}^2+171\right)}{\pi 
   \left(\mathcal{N}^2-4\right)^2 }\,\frac{\alpha^{3}}{\tau^{11}}\nonumber \\
   &&-\frac{26 k \left(6
   \mathcal{N}^6-257 \mathcal{N}^4+3668 \mathcal{N}^2-16569\right)}{\pi 
   \left(\mathcal{N}^2-4\right)^3}\,\frac{\alpha^{4}}{\tau^{14}}\nonumber \\
   &&+\frac{192 k \left(11 \mathcal{N}^8-751
   \mathcal{N}^6+19206 \mathcal{N}^4-212291 \mathcal{N}^2+830241\right)}{\pi 
   \left(\mathcal{N}^2-4\right)^4 }\,\frac{\alpha^{5}}{\tau^{17}}+\mc O(\alpha ^6),\\
 \mc J_{4} &=& \frac{3   k \left(\mathcal{N}^2-9\right)}{7 \pi  \left(\mathcal{N}^2-4\right) }\,\frac{\alpha}{\tau^{7}}
 -\frac{3  k \left(\mathcal{N}^4-28 \mathcal{N}^2+171\right)}{\pi 
   \left(\mathcal{N}^2-4\right)^2 }\,\frac{\alpha^{2}}{\tau^{10}}\nonumber \\
   &&+\frac{8  k \left(6 \mathcal{N}^6-257
   \mathcal{N}^4+3668 \mathcal{N}^2-16569\right)}{\pi  \left(\mathcal{N}^2-4\right)^3 }\,\frac{\alpha^{3}}{\tau^{13}}\nonumber \\
   &&-\frac{60  k \left(11 \mathcal{N}^8-751 \mathcal{N}^6+19206
   \mathcal{N}^4-212291 \mathcal{N}^2+830241\right)}{\pi  \left(\mathcal{N}^2-4\right)^4
   }\,\frac{\alpha^{4}}{\tau^{16}}+\mc O(\alpha ^5), \\
\mc J_{6} &=& \frac{11 \alpha ^2}{\left(\mathcal{N}^2-4\right)^2 \tau ^{12}}-\frac{44
   \alpha ^3 \left(12 \mathcal{N}^2-403\right)}{3
   \left(\left(\mathcal{N}^2-4\right)^3 \tau ^{15}\right)}+\frac{44
   \alpha ^4 \left(78 \mathcal{N}^4-4764
   \mathcal{N}^2+83933\right)}{\left(\mathcal{N}^2-4\right)^4 \tau
   ^{18}}\nonumber \\
   &&-\frac{4400 \alpha ^5 \left(14 \mathcal{N}^6-1363
   \mathcal{N}^4+49472
   \mathcal{N}^2-649951\right)}{\left(\mathcal{N}^2-4\right)^5 \tau
   ^{21}}+\mc O(\alpha ^6).
 \ea
 The expansions trivialise at $\N=3$ as it should be.
 The condition $\partial_{\alpha} \mc L=\partial_{\tau} \mc J_{4}$ is indeed satisfied and implies that
 \be
 \label{eq:J2J4constraint}
 \mc J_{4,n}(\N) = -\frac{n+1}{3\,n+4}\,\mc L_{n+1}(\N),\qquad n\ge 1\,.
 \ee 
 A partition function can
 be defined integrating the relation
 \be
 \mc L(\alpha, \tau) = \frac{1}{4\,\pi^{2}\,i}\,\frac{\partial}{\partial\tau}\log Z_{\N}(\alpha, \tau),
 \ee
 that simply implies
 \be
 \label{eq:Zexp}
\log Z_{\N}(\alpha, \tau) = \sum_{n=0}^{\infty} b_{n}(\N)\,\frac{\alpha^{n}}{\tau^{3n+1}}, \qquad
b_{n}(\N) = -\frac{4\,\pi^{2}\,\tau\,i}{3n+1}\,\mc L_{n}(\N).
 \ee


\section{CFT calculation of the partition function at $\N=0$ and $1$}
\label{sec:CFTtest}

Our perturbative calculation provides the partition function with its explicit dependence on $\N$ 
order by order in $\alpha/\tau^{3}$. This allows to perform an important check 
of $AdS_{3}/CFT_{2}$ duality. The generic $\N$ result can be uplift to the theory 
with symmetry $\hs\oplus\hs$ where $\lambda=\N$. 
According to \cite{Gaberdiel:2010pz}, the dual conformal theory has chiral symmetry 
algebra $\mc W_{\infty}[\lambda]$. This infinite dimensional algebra 
admits a free boson (fermion) realisations at $\lambda=1\,(0)$. 
Thus, the partition function evaluated at these special values of $\N$ 
can be matched to a CFT calculation in such simple theories. This calculation has been
performed in \cite{Kraus:2011ds} for the spin-3 higher spin black hole. Here, we shall 
apply the same methods to our case. 

\subsection{Free bosons, $\N=1$}
The partition function at $\N=1$ reads
\ba
\label{eq:N1logZ}
\lefteqn{\log Z_{\N=1}(\tau,\alpha) = \frac{i\,\pi\,k}{2}\,\bigg[} && \nonumber \\
&&  \frac{1}{\tau }+\frac{32 \alpha ^2}{7 \tau ^7}-\frac{128 \alpha ^3}{\tau ^{10}}+\frac{70144
   \alpha ^4}{9 \tau ^{13}}-\frac{20365312 \alpha ^5}{27 \tau ^{16}}+\frac{2899976192 \alpha
   ^6}{27 \tau ^{19}}-\frac{5142223683584 \alpha ^7}{243 \tau ^{22}}\nonumber \\
   &&+\frac{28160000624820224
   \alpha ^8}{5103 \tau ^{25}}-\frac{149140912760422400 \alpha ^9}{81 \tau
   ^{28}}+\frac{35151331424584119353344 \alpha ^{10}}{45927 \tau
   ^{31}}\nonumber \\
   &&-\frac{53447811794486723086385152 \alpha ^{11}}{137781 \tau
   ^{34}}+\frac{32448962211691325404330590208 \alpha ^{12}}{137781 \tau
   ^{37}}\nonumber \\
   &&-\frac{209252647957707775596720186982400 \alpha ^{13}}{1240029 \tau
   ^{40}}+\frac{5766272854912622868593398402112290816 \alpha ^{14}}{40920957 \tau
   ^{43}}\nonumber \\
   &&-\frac{5551159515945111623206530290764658769920 \alpha ^{15}}{40920957 \tau
   ^{46}}\nonumber \\
   &&+\frac{4993611566073029965492725902288304342040576 \alpha ^{16}}{33480783 \tau
   ^{49}}\nonumber \\
   &&-\frac{18675117272453224775576949583343472730854580551680 \alpha ^{17}}{100542791349
   \tau ^{52}}\nonumber \\
   &&+\frac{670564221083893280300893857178854185446940616949760 \alpha
   ^{18}}{2578020291 \tau ^{55}}+\mc O(\alpha ^{19})\bigg].
   \ea
The theory of $D$ free complex bosons $\varphi^{i}$, $i=1, \dots D$, has  $\mc W_{\infty}[1]$ symmetry and  central charge
$c = 2\,D$ \cite{Hull:1991sa,Hull:1993kf}. The algebra can be recast in the linear form $\mc W_{\infty}^{\rm PRS}$
with quadratic currents~\cite{Pope:1989sr}. 
The first cases are \cite{Bakas:1990ry} (see also 
the discussion in \cite{Gaberdiel:2013jpa})
\ba
W^{(2)} \equiv T &=& -:\partial\varphi^{j}\partial\overline\varphi^{j}:\,,\\
W^{(3)} &=& -2\,:(\partial\varphi^{j}\partial^{2}\overline\varphi^{j}-\partial^{2}\varphi^{j}\partial\overline\varphi^{j}):\, , \\
W^{(4)} &=& -\frac{16}{5}\,:(
\partial\varphi^{j}\partial^{3}\overline\varphi^{j}
-3\partial^{2}\varphi^{j}\partial^{2}\overline\varphi^{j}
+\partial^{3}\varphi^{j}\partial\overline\varphi^{j}
):\, .
\ea
The partition function evaluated with the insertion of the zero mode of $W^{(4)}$ is~\footnote{We 
remove a normal ordering constant in the Virasoro charge since it is not relevant to the high temperature expansion.}
\be
\widetilde Z_{\mc N=1}(\tau, \alpha) = \mbox{Tr}\bigg[e^{4\pi^{2}
i\,(\tau\, W^{(2)}_{0}+\alpha\, W^{(4)}_{0})}\bigg]
\ee
Adapting the calculation of  \cite{Kraus:2011ds}, this can be written in the high temperature limit $\tau\to 0$
\be
\log\widetilde Z_{\mc N=1}(\tau, \alpha) = -\frac{6}{\tau\,\pi^{2}}\int_{0}^{\infty}\log\left(
1-e^{-x+a\,\frac{\alpha}{\tau^{3}}\,x^{3}}\right),\qquad a=-\frac{5}{3\,\pi^{2}}.
\ee
The integral can be evaluated by the methods described in \cite{Beccaria:2013dua} and we obtain the exact expansion
\ba
\log\widetilde Z_{\mc N=1}(\tau, \alpha) &=& 12\,\sum_{n=0}^{\infty}\left(\frac{20}{3}\right)^{n}\,\frac{B_{2n+2}\,\Gamma(3n+1)}{\Gamma(n+1)\,\Gamma(2n+3)}\,\frac{\alpha^{n}}{\tau^{3n+1}} =  \\
&=& \frac{1}{\tau }-\frac{2 \alpha }{3 \tau ^4}+\frac{400 \alpha ^2}{63 \tau ^7}-\frac{1600 \alpha
   ^3}{9 \tau ^{10}}+\frac{800000 \alpha ^4}{81 \tau ^{13}}-\frac{221120000 \alpha ^5}{243 \tau
   ^{16}}+\mc O(\alpha^{6}),\nonumber 
\ea
where $B_{n}$ are Bernoulli numbers.
The current $W^{4}$ is not Virasoro primary, but can be made primary by the addition of an operator whose contribution
to the charge is proportional to $[W^{(2)}_{0}]^{2}$. This contribution can be taken into account by computing
the quantity
\be
\label{eq:op1}
\log\left[
\exp\left(\frac{2}{3}\,\alpha\,\frac{\partial^{2}}{\partial\tau^{2}}\right)\,\widetilde Z_{\mc N=1}(\tau, \alpha)
\right].
\ee
Expanding in powers of $\alpha$ we obtain the series
\ba
&& \frac{1}{\tau }+\frac{4 \alpha }{3 \tau ^3}+\alpha ^2 \left(\frac{32}{7 \tau ^7}-\frac{16}{9
   \tau ^6}+\frac{16}{3 \tau ^5}\right)+\alpha ^3 \left(-\frac{128}{\tau ^{10}}+\frac{14080}{81
   \tau ^9}-\frac{128}{3 \tau ^8}+\frac{320}{9 \tau ^7}\right)\nonumber \\
   &&+\alpha ^4 \left(\frac{70144}{9
   \tau ^{13}}-\frac{797696}{81 \tau ^{12}}+\frac{48128}{9 \tau ^{11}}-\frac{24064}{27 \tau
   ^{10}}+\frac{8960}{27 \tau ^9}\right)+\mc O(\alpha ^5).
\ea
Going to the limit $\tau, \alpha\to 0$ with fixed $\alpha/\tau^{3}$, most terms are subleading and the surviving contributions (the most singular terms for $\tau\to 0$) agree with (\ref{eq:N1logZ}). We have 
checked the agreement up to the available order $\mc O(\alpha^{18})$.
For the following discussion, it is convenient to introduce a notation for the operation in (\ref{eq:op1})
followed by the high temperature limit. To this aim, let us consider the partition function at a certain $\N$
(we shall be interested in the cases $\N=0,1,\infty$). It has the structure $\log Z(\tau, \alpha) = \frac{1}{\tau}\,f\left(\frac{\alpha}{\tau^{3}}\right)$. For such a function, we introduce the operator $\mc S_{\gamma}$ by the definition
\be
\mc S_{\gamma}[\log Z(\tau,\alpha)] = \frac{1}{\tau}\mathop{\lim_{\tau\to 0}}_{\alpha/\tau^{3}\, \rm fixed} \tau\log\left[
e^{\gamma\,\alpha\,\frac{\partial^{2}}{\partial\tau^{2}}}\,Z(\tau,\alpha)
\right].
\ee
This operator returns a new partition function, {\em i.e.} a new series of the form 
$\frac{1}{\tau}\,\widetilde f\left(\frac{\alpha}{\tau^{3}}\right)$, with a transformed $f\to \widetilde f$. The result
at $\N=1$ is thus 
\be
\label{eq:bosonresult}
\log Z_{\N=1}(\tau, \alpha) = \mc S_{\frac{2}{3}}[\log \widetilde Z_{\N=1}(\tau,\alpha)].
\ee

\subsection{Free fermions, $\N=0$}
The partition function at $\N=0$ reads
\ba
\label{eq:N0logZ}
\lefteqn{\log Z_{\N=0}(\tau,\alpha) = \frac{i\,\pi\,k}{2}\,\bigg[} && \nonumber \\
&& 
\frac{1}{\tau }+\frac{27 \alpha ^2}{7 \tau ^7}-\frac{171 \alpha ^3}{2 \tau ^{10}}+\frac{16569
   \alpha ^4}{4 \tau ^{13}}-\frac{2490723 \alpha ^5}{8 \tau ^{16}}+\frac{545057541 \alpha
   ^6}{16 \tau ^{19}}-\frac{163877633451 \alpha ^7}{32 \tau ^{22}}\nonumber \\
   &&+\frac{454476030924519 \alpha
   ^8}{448 \tau ^{25}}-\frac{32814851841987075 \alpha ^9}{128 \tau
   ^{28}}+\frac{20620715931205458957 \alpha ^{10}}{256 \tau
   ^{31}}\nonumber \\
   &&-\frac{110435225942799025314093 \alpha ^{11}}{3584 \tau
   ^{34}}+\frac{14442756275364844352986377 \alpha ^{12}}{1024 \tau
   ^{37}}\nonumber \\
   &&-\frac{15591545452041690158496864675 \alpha ^{13}}{2048 \tau
   ^{40}}+\frac{1509460994349551806229081175269841 \alpha ^{14}}{315392 \tau
   ^{43}}\nonumber \\
   &&-\frac{312407453050840141214902163578176345 \alpha ^{15}}{90112 \tau
   ^{46}}\nonumber \\
   &&+\frac{46971962777346668301404387561533369713 \alpha ^{16}}{16384 \tau
   ^{49}}\nonumber \\
   &&-\frac{88053553875786370110480192173535873605066235 \alpha ^{17}}{32800768 \tau
   ^{52}}\nonumber \\
   &&+\frac{2037065943846643106948011139763108699311974335 \alpha ^{18}}{720896 \tau
   ^{55}}+\mc O(\alpha ^{19})\bigg].
   \ea
The theory of $D$ free complex fermions has $\mc W_{1+\infty}$ symmetry and central charge $c = D$ 
\cite{Bergshoeff:1991dz}. The $\mc W_{1+\infty}$ algebra has currents with spin $1,2,3,\dots$ and leads to 
$\mc W_{\infty}[0]$ by imposing a constraint that eliminates the spin-1 current. In the black hole with spin-3 chemical 
potential this was achieved by introducing a further chemical potential for the spin-1 charge and requiring that charge to vanish.
In the spin-4 black hole solution considered here, this is not necessary since parity automatically sets it to zero.
In close analogy to the previous bosonic case, and using the same notation, the high temperature limit $\tau\to 0$
of the partition function with the insertion of the non-primary spin-4 charge is now
\be
\log\widetilde Z_{\mc N=0}(\tau, \alpha) = \frac{12}{\tau\,\pi^{2}}\int_{0}^{\infty}\log\left(
1+e^{-x+a\,\frac{\alpha}{\tau^{3}}\,x^{3}}\right),\qquad a=-\frac{5}{4\,\pi^{2}}.
\ee
Again, the integral can be evaluated and we obtain the exact expansion
\ba
\log\widetilde Z_{\mc N=0}(\tau, \alpha) &=& 12\,\sum_{n=0}^{\infty}\left(\frac{5}{4}\right)^{n}\,\frac{
(2^{2n+1}-1)\,B_{2n+2}\,\Gamma(3n+1)}{\Gamma(n+1)\,\Gamma(2n+3)}\,\frac{\alpha^{n}}{\tau^{3n+1}} =  \\
&=&
\frac{1}{\tau }-\frac{7 \alpha }{8 \tau ^4}+\frac{775 \alpha ^2}{112 \tau ^7}-\frac{9525 \alpha
   ^3}{64 \tau ^{10}}+\frac{1596875 \alpha ^4}{256 \tau ^{13}}-\frac{884048125 \alpha ^5}{2048
   \tau ^{16}}+\mc O(\alpha^{6}).\nonumber 
\ea
Even in this case, the spin-4 current is not Virasoro primary, but it can be made so
by modifying the charge with a term $\sim [W^{(2)}_{0}]^{2}$. In full details, the partition function is 
recovered by the formula
\be
\label{eq:fermionresult}
\log Z_{\N=0}(\tau, \alpha) = \mc S_{\frac{7}{8}}[\log \widetilde Z_{\N=0}(\tau,\alpha)].
\ee
that we have checked up to the order $\mc O(\alpha^{18})$. This is the same as (\ref{eq:bosonresult})
with the simple replacement $2/3\to 7/8$ to take into account the explicit form of the primary spin-4 current in the 
free fermionic theory.
%

\section{The partition function at $\N=\infty$}
\label{sec:Ninfty}

By analogy with the spin-3 case, it is interesting
to evaluate the partition function in the $\N\to\infty$ limit. 
This gives the following series in the ratio $\alpha/\tau^{3}$
  \ba
\lefteqn{\log Z_{\N=\infty}(\tau,\alpha) = \frac{i\,\pi\,k}{2}\,\bigg[} && \nonumber \\
&&     \frac{1}{\tau }+\frac{12 \alpha ^2}{7 \tau ^7}-\frac{8 \alpha ^3}{\tau ^{10}}+\frac{96 \alpha
   ^4}{\tau ^{13}}-\frac{1056 \alpha ^5}{\tau ^{16}}+\frac{14016 \alpha ^6}{\tau
   ^{19}}-\frac{196608 \alpha ^7}{\tau ^{22}}+\frac{2949888 \alpha ^8}{\tau
   ^{25}}-\frac{323980800 \alpha ^9}{7 \tau ^{28}}\nonumber \\
   &&+\frac{754486272 \alpha ^{10}}{\tau
   ^{31}}-\frac{12682616832 \alpha ^{11}}{\tau ^{34}}+\frac{218770444288 \alpha ^{12}}{\tau
   ^{37}}-\frac{3857074176000 \alpha ^{13}}{\tau ^{40}}\nonumber \\
   &&+\frac{69291997052928 \alpha ^{14}}{\tau
   ^{43}}-\frac{1265276167618560 \alpha ^{15}}{\tau ^{46}}+\frac{164054123598249984 \alpha
   ^{16}}{7 \tau ^{49}}\nonumber \\
   &&-\frac{439613746473861120 \alpha ^{17}}{\tau
   ^{52}}+\frac{8339276221242408960 \alpha ^{18}}{\tau ^{55}}+\mc O(\alpha ^{19})\bigg].
\ea
In this case, we did not succeed in finding a resummation for the above expansion. Therefore,
we cannot identify analytically a critical 
value for $\alpha/\tau^{3}$ analogous to  (\ref{eq:spin3-alphac}). Nevertheless, inspired by the simple
relations (\ref{eq:bosonresult}) and (\ref{eq:fermionresult}), we tried to see if a  generalisation in terms of 
$\mc S_{\gamma}$ with a suitable $\gamma$ happens to 
simplify the expansion of the partition function. Remarkably, one finds that $\gamma=1$ does the job. Indeed, 
\ba
&& \mc S_{1}\left[\frac{2}{i\,\pi\,k}\log Z_{\N=\infty}(\tau, \alpha)\right] = F(\tau,\alpha)=\nonumber \\
&& \qquad \frac{1}{\tau }+\frac{\alpha }{\tau ^4}+\frac{40 \alpha ^2}{7 \tau ^7}+\frac{40 \alpha ^3}{\tau ^{10}}+\frac{400
   \alpha ^4}{\tau ^{13}}+\frac{4000 \alpha ^5}{\tau ^{16}}+\frac{48000 \alpha ^6}{\tau ^{19}}+\frac{560000
   \alpha ^7}{\tau ^{22}}+\frac{7360000 \alpha ^8}{\tau ^{25}}\nonumber \\
   &&+\frac{93600000 \alpha ^9}{\tau
   ^{28}}+\frac{1299200000 \alpha ^{10}}{\tau ^{31}}+\frac{17459200000 \alpha ^{11}}{\tau
   ^{34}}+\frac{251328000000 \alpha ^{12}}{\tau ^{37}}\nonumber \\
   &&+\frac{3509376000000 \alpha ^{13}}{\tau
   ^{40}}+\frac{51850240000000 \alpha ^{14}}{\tau ^{43}}+\frac{744691200000000 \alpha ^{15}}{\tau
   ^{46}}+\mc O(\alpha ^{16}).
\ea
The series in the r.h.s. is simpler than the expansion of the partition function. Indeed, with some work, 
it can be resummed in closed form as follows
\be
F(\tau,\alpha) = \frac{\tau^{2}}{12\,\alpha}\left[
{}_{5}F_{4}\left(\left. 
\begin{array}{c}
-\frac{1}{3}, \ -\frac{1}{6},\ \frac{1}{6},\ \frac{1}{3},\ \frac{1}{2} \\
\frac{1}{5},\ \frac{2}{5},\ \frac{3}{5},\ \frac{4}{5}
\end{array}
\right| \left(\frac{432\alpha}{25\tau^{3}}\right)^{2}\right)-1
\right]+\frac{1}{\tau}\,
{}_{5}F_{4}\left(\left. 
\begin{array}{c}
\frac{1}{6}, \ \frac{1}{3},\ \frac{1}{2},\ \frac{2}{3},\ \frac{5}{6} \\
\frac{3}{5},\ \frac{4}{5},\ \frac{6}{5},\ \frac{7}{5}
\end{array}
\right| \left(\frac{432\alpha}{25\tau^{3}}\right)^{2}\right).
\ee
This function is the generating function of $\log Z_{\N=\infty}$ in the sense that the partition function can be 
reconstructed by 
\be
\label{eq:Ninftyresult}
\log Z_{\N=\infty}(\tau,\alpha) = \frac{i\,\pi\,k}{2}\,\mc S_{-1}\left[F(\tau,\alpha)\right].
\ee
In App.~(\ref{app:expansion}), we collect the extended expansion of $\log Z_{\N=\infty}(\tau,\alpha)$
up to order $\mc O(\alpha^{33})$ computed from the generating function. 

As a final important remark, 
we emphasise that the critical point of $F(\tau,\alpha)$ does not coincide with the radius of convergence of 
$\log Z_{\N=\infty}(\tau,\alpha)$ as one could naively guess. The reason is that the $\mc S_{\gamma}$ operation is 
highly non-trivial and generically changes the radius of convergence of the series to which it is applied. This statement can be easily checked by working out simple model functions for the partition function and analysing the associated series expansions.
   
\section{The BTZ critical temperature of higher spin black holes}
\label{sec:CriticalTemperature}

\subsection{The spin-3 case}

\begin{figure}[htbp]
	\centering
	\includegraphics[scale=0.5]{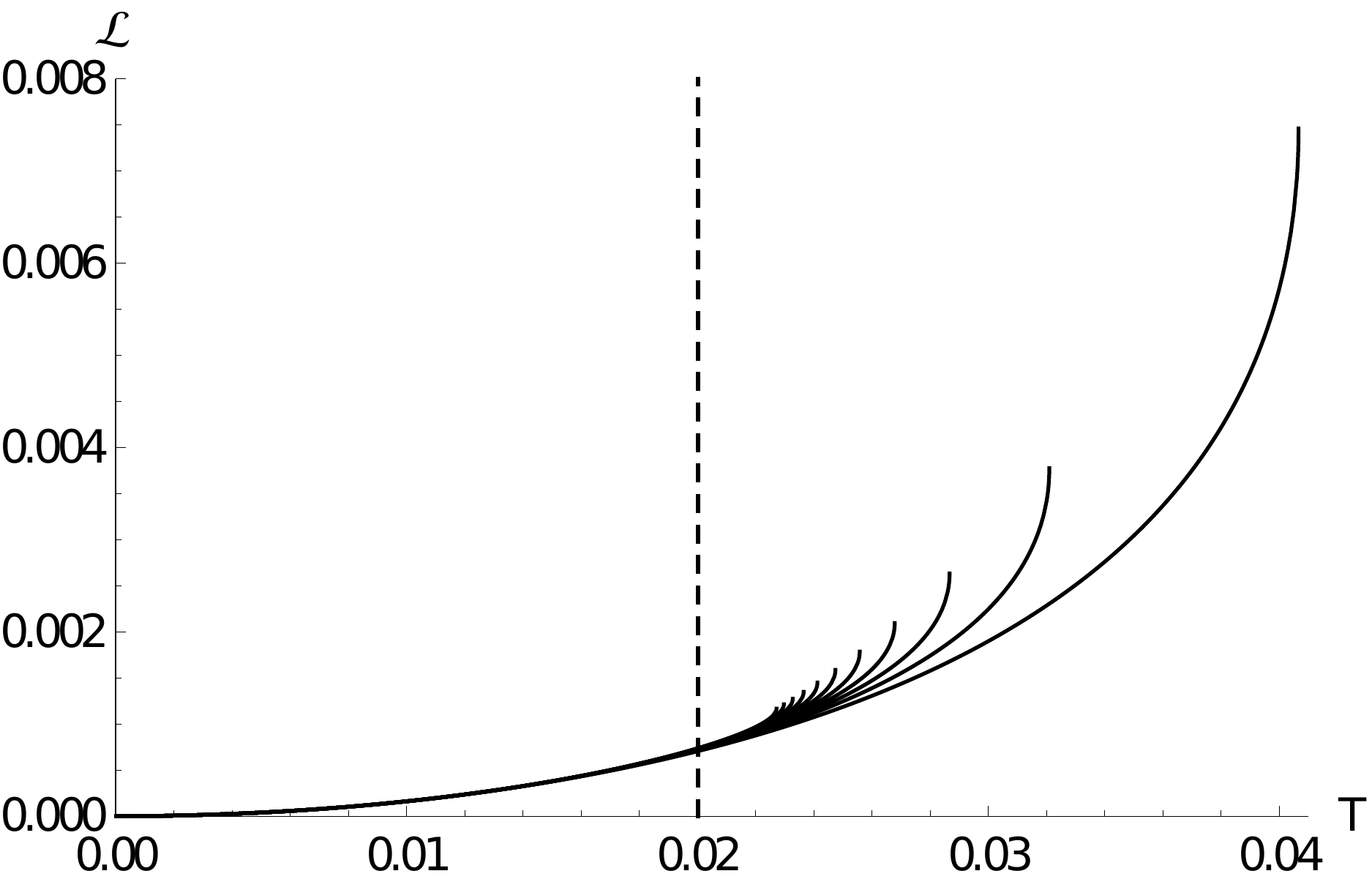}
	\caption{Spin-3 black hole. BTZ branch for the function $\mc L(T)$  at $\N=3, 4, \dots, 13$, from right to left. The dashed vertical line is placed at the expected asymptotic value $T_{\rm BTZ}(\N=\infty) = \frac{9}{64\,\pi\,\sqrt 5}$.
}
	\label{fig:Spin3-Ta}
\end{figure}

\begin{figure}[htbp]
	\centering
	\includegraphics[scale=0.5]{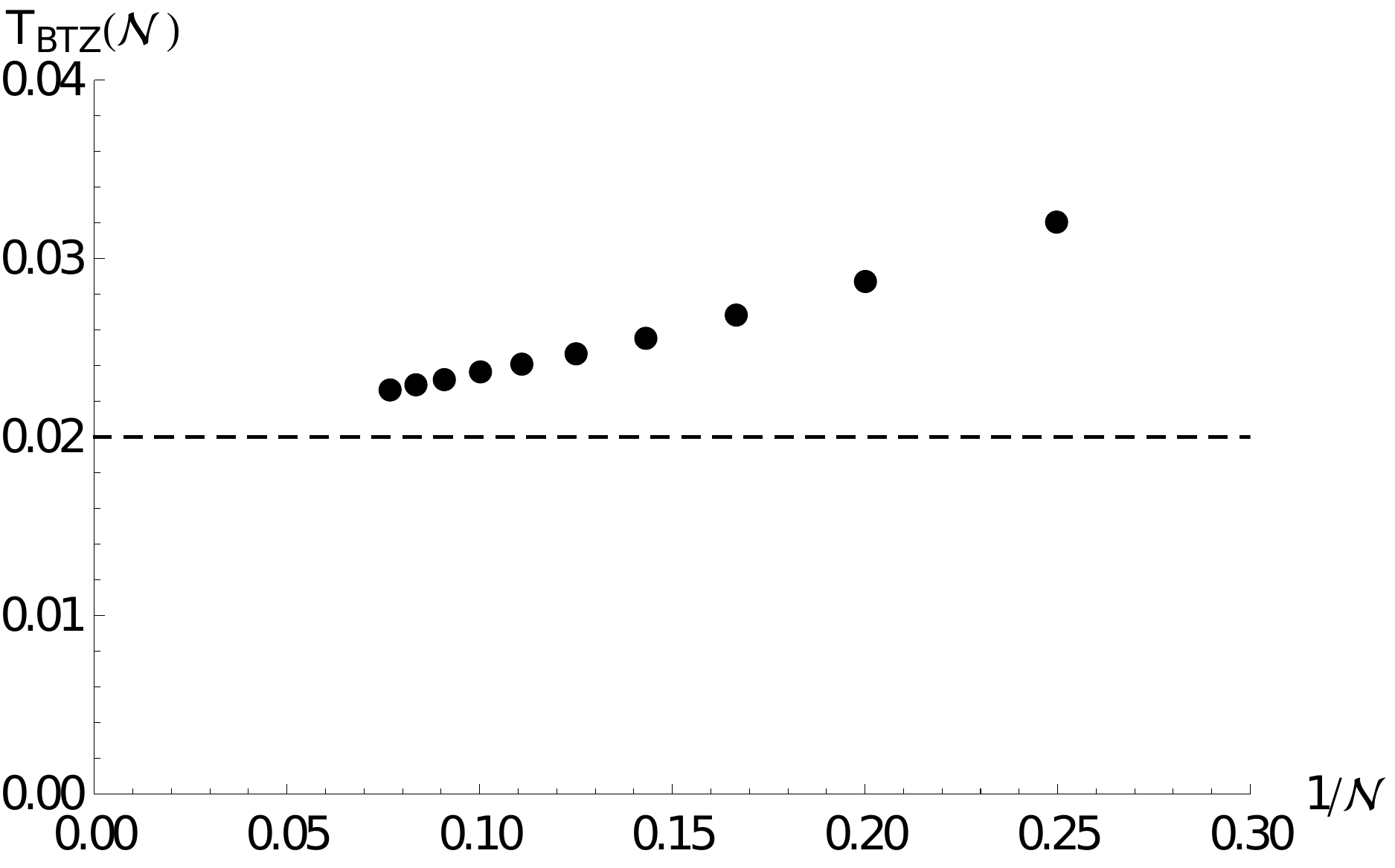}	
	\caption{Spin-3 black hole. Critical BTZ temperature at $\N=3, 4, \dots, 13$, from right to left. The dashed vertical line is placed at the expected asymptotic value $T_{\rm BTZ}(\N=\infty) = \frac{9}{64\,\pi\,\sqrt 5}$.
}
	\label{fig:Spin3-Tb}
\end{figure}

As discussed in \cite{David:2012iu} for $\N=3$, the holonomy equations admits several branches of solutions whose physical meaning is quite interesting. In particular, there is a (BTZ) branch which is associated with the perturbative expansion in $\alpha$ . The BTZ branch can be studied at fixed chemical potential $\mu$ as a function of the temperature $T$ appearing in 
\be
\tau = \frac{i}{2\,\pi\,T},\qquad \alpha = \overline\tau\,\mu = -\frac{i}{2\,\pi\,T}\,\mu.
\ee
It is found that the BTZ branch exists up to a critical temperature $T_{BTZ}$. The extension of these analysis to the 
spin-3 black hole in $\mk{sl}(\N)$ can be done analytically at low $\N$, but requires some numerics already for moderately large $\N$ because of the complexity of the holonomy conditions. In the following, we shall set $\mu=1$ without loosing generality since the holonomy equations determine a critical value of the product $\mu\,T$. Also, 
there is symmetry between $\mu\to -\mu$. We computed accurately the BTZ branch by starting at $T=0$ and increasing it in adaptive steps $\Delta T$ using at each step the solution computed at the previous step. The numerics has been evaluated with typically $60$ digits. From  (\ref{eq:spin3-alphac}), we expect to find
\be
T_{\rm BTZ}(\N=\infty) = \frac{9}{64\,\pi\,\sqrt 5}.
\ee
In \figref{Spin3-Ta}, we show that BTZ branch for the function $\mc L(T)$ computed at $\N=3, 4, \dots, 13$. Each curve
starts from $\mc L(0)=0$ and increases along the perturbative expansion up to a point where the curve develops a vertical tangent. This is the point where a second  upper branch of the holonomy conditions, with $\mc L(0)\neq 0$, merges
with the BTZ branch. The critical value $T_{\rm BTZ}(\N)$ seems to converge to the expected value which is marked with a vertical dashed line. Their explicit expression is given in the following table
\be
\begin{array}{|l|l|}
\hline
\N & T_{\rm BTZ}(\N) \\
\hline
3 & 0.0406591166737577094712068073810 \\
4 & 0.0320927194886359815492098778510 \\
5 & 0.0286647278507052620641709242510 \\
6 & 0.0267788304143913750916077956210 \\
7 & 0.0255762577941300149772795461610 \\
8 & 0.0247396632171781771847834319210 \\
\hline
\end{array}\quad
\begin{array}{|l|l|}
\hline
\N & T_{\rm BTZ}(\N) \\
\hline
9 & 0.0241229488371714796928574597010 \\
10 & 0.0236490377011931115967660930110 \\
11 & 0.0232732681263221309465627961310 \\
12 & 0.0229679153999352997373903457810 \\
13 & 0.0227148331154236883304377338910 \\ & \\ 
\hline
\end{array}
\ee
The convergence is illustrated graphically in \figref{Spin3-Tb} where the dashed line is again the predicted asymptotic value. We extrapolated the sequence $\{T_{\rm BTZ}(\N)\}$ at $\N=\infty$ using the Bulirsch-Stoer (BST) algorithm 
discussed in App.~(\ref{app:BST}) with the result
\be
\left. T_{\rm BTZ}(\infty)\right|_{\rm Bulirsch-Stoer} = 0.0200(1).
\ee
This compares perfectly with the analytical value $\frac{9}{64\,\pi\,\sqrt 5}=0.0200183\dots$. In summary, this analysis 
has proved that there is a critical value of the temperature for each $\N$ and that the sequence of values $T_{\rm BTZ}(\N)$ converges to the analytical prediction.  As $\N$ increases, the curves collapse to a smooth curve of finite length. This is 
consistent with the $\N=\infty$ analytical result. Indeed, the inflection of the BTZ branch disappears in this limit and the
following derivatives are finite
\ba
\left. \frac{d\mc L}{dT}\right|_{T_{\rm BTZ}(\infty)} &=& 
\frac{9}{64\,\sqrt 5}\,\hyper{\frac{1}{4}, \ \frac{1}{2}, \ \frac{3}{4}}{\frac{4}{3}, \ \frac{5}{3}}{1}+\frac{243}{10240\,\sqrt 5}\,\hyper{\frac{5}{4},\ \frac{3}{2},\ \frac{7}{4}}{\frac{7}{3},\ \frac{8}{3}}{1} = \frac{\sqrt 5}{20},\\
\left. \frac{d\mc W}{dT}\right|_{T_{\rm BTZ}(\infty)} &=& 
\frac{243}{20480\,\sqrt 5}\,\hyper{\frac{5}{4},\ \frac{3}{2},\ \frac{7}{4}}{\frac{7}{3},\ \frac{8}{3}}{1} = \frac{\sqrt 5}{100},
\ea
with a similar result for 
$\lim_{\N\to \infty}\N^{2}\left. \frac{d\mc J_{4}}{dT}\right|_{T_{\rm BTZ}(\N)} = \frac{21\,\pi\,\sqrt 5}{3200}$.

\subsection{The spin-4 case}

\begin{figure}[htbp]
	\centering
	\includegraphics[scale=0.6]{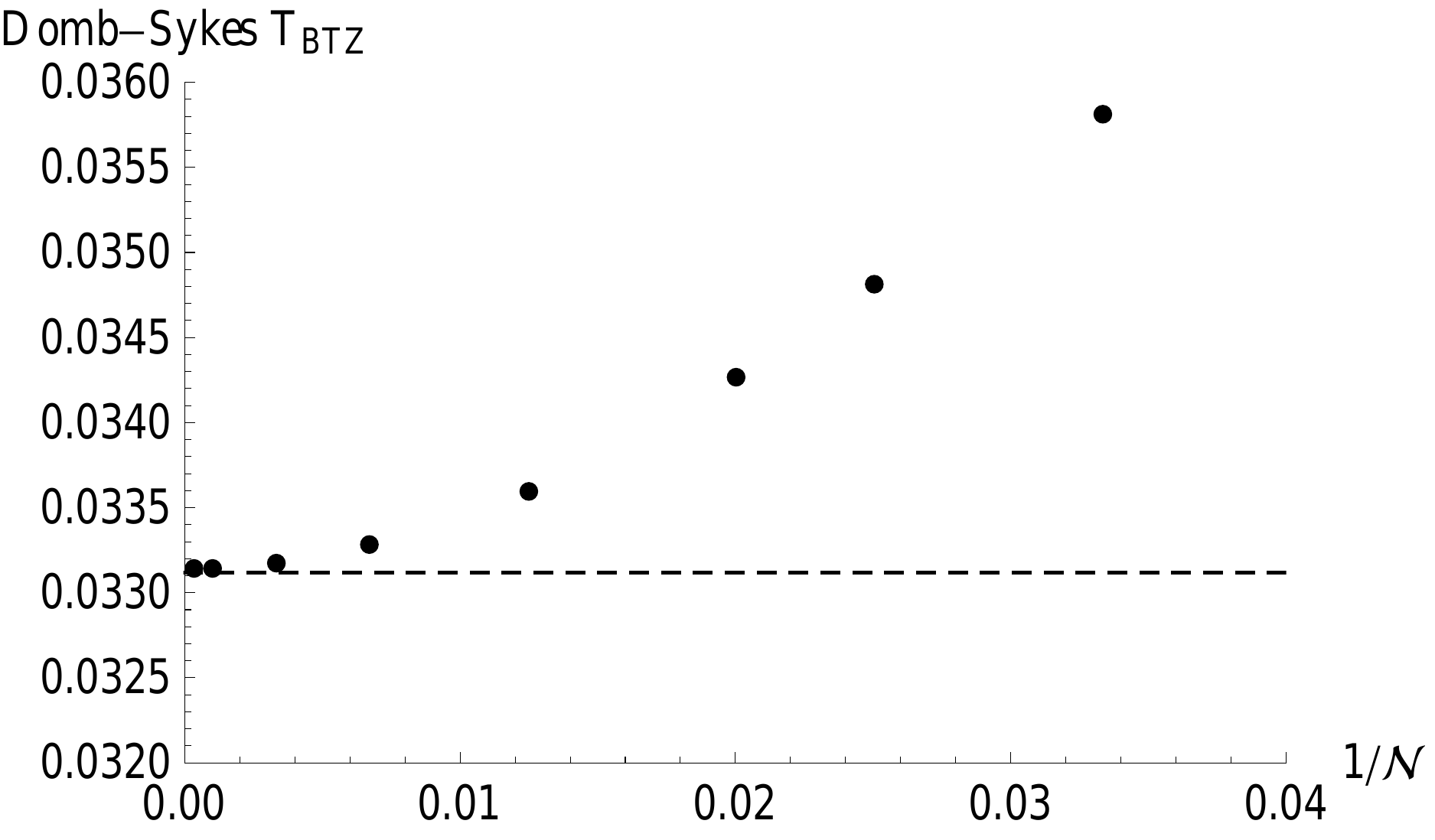}	
	\caption{Spin-4 black hole. Domb-Sykes estimate of the BTZ temperature versus $1/\N$. Data points correspond
	to $\N = 30, 40, 50, 80, 150, 300, 1000, 3000$ from right to left. These values are replaced in the perturbative series 
	for the partition function and an estimate of the radius of convergence is done as explained in the text.
	The dashed line is at $T_{\rm DS}$ and guides the eye to show convergence.}
	\label{fig:Domb}
\end{figure}

In the spin-4 case, we expect the same qualitative picture to be valid for the BTZ branch. However, 
there is no available closed form for the partition function not even in the large $\N$ limit. Nevertheless,
we can take the perturbative expansion of $\log Z$ at a certain $\N$ and estimate the critical temperature by 
identifying it with the radius of convergence of the perturbative series. This is done by plotting a Domb-Sykes plot \cite{Domb-Sykes} where the ratio of subsequent coefficients of the power series is studied as a function of the inverse of their index. If a linear trend is asymptotically observed, then its intercept  determines the convergence radius. 
Doing this exercise, and taking $|\mu|=1$, we obtain \figref{Domb}. The dashed line is placed at $T_{\rm DS}$
which is obtained from the Domb-Sykes extrapolation of the remarkably  long series at $\N=\infty$
reported in App.~(\ref{app:expansion}). It can be determined rather accurately and reads
\be
T_{\rm DS}=0.033184(5).
\ee

\begin{figure}[htbp]
	\centering
	\includegraphics[scale=0.5]{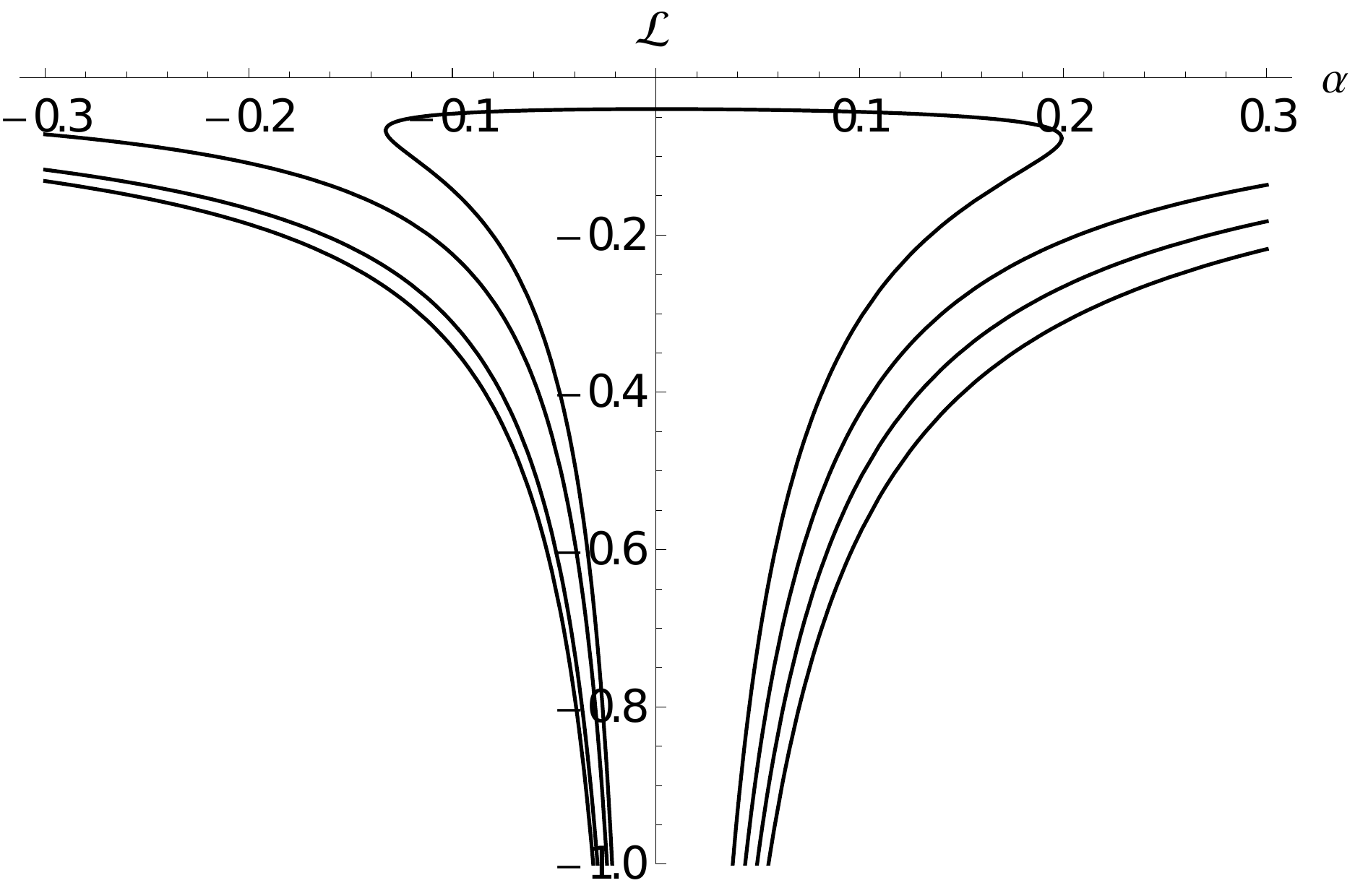}	
	\caption{Spin-4 black hole. Some branches of  the function $\mc L(\alpha)$  at $\N=5$, $\tau=1$, including the 
	BTZ branch which is the upper curve.
	There are two asymmetric singular values where the curve has a vertical tangent.}
	\label{fig:Spin4-N5}
\end{figure}

The exact determination of the critical temperature at fixed moderate $\N$ can be done as in the previous case
of the spin-3 black hole. There is however a remarkable difference already noted in \cite{Chen:2012ba}.
In the spin-3 case, the partition function 
is an even function of $\alpha/\tau^{2}=-2\,\pi\,i\,\mu\,T$. Hence, the sign of $\mu$ is irrelevant. Here, 
the partition function is a function of $\alpha/\tau^{3}=4\,\pi^{2}\,\mu\,T^{2}$. So, changing the sign of $\mu$
allows to explore the partition function at different signs of $\alpha/\tau^{3}$. Now, the convergence radius of the partition function is the value of $|\alpha/\tau^{3}|$ at the singularity which is nearest to the origin. It turns out that there are two 
almost symmetric singularities according to the sign of $\alpha/\tau^{3}$. 
This is illustrated in \figref{Spin4-N5} where we plot the branches of the function $\mc L(\alpha/\tau^{3})$ at $\tau=1$
and $\N=5$. The BTZ branch is the upper curve and has the property of being finite at $\alpha=0$, according to the perturbative expansion that we derived. It exists up to a critical value that is different on the two sides of the figure. The convergence radius of the perturbative series is clearly the one on the left, that is the nearest to the origin.
In conclusion, a proper choice of the sign of $\mu$ is required if we want to identify the convergence ratio with the $\N\to\infty$ limit of the critical temperatures of the BTZ branch. We have analysed this issue and found that $\mu<0$ is correct. Thus, in the following we shall take $\mu=-1$. Of course, the picture at $\mu>0$ is similar, but without coincidence between 
$T_{\rm BTZ}(\infty)$ and the partition function convergence radius.

\begin{figure}[htbp]
	\centering
	\includegraphics[scale=0.5]{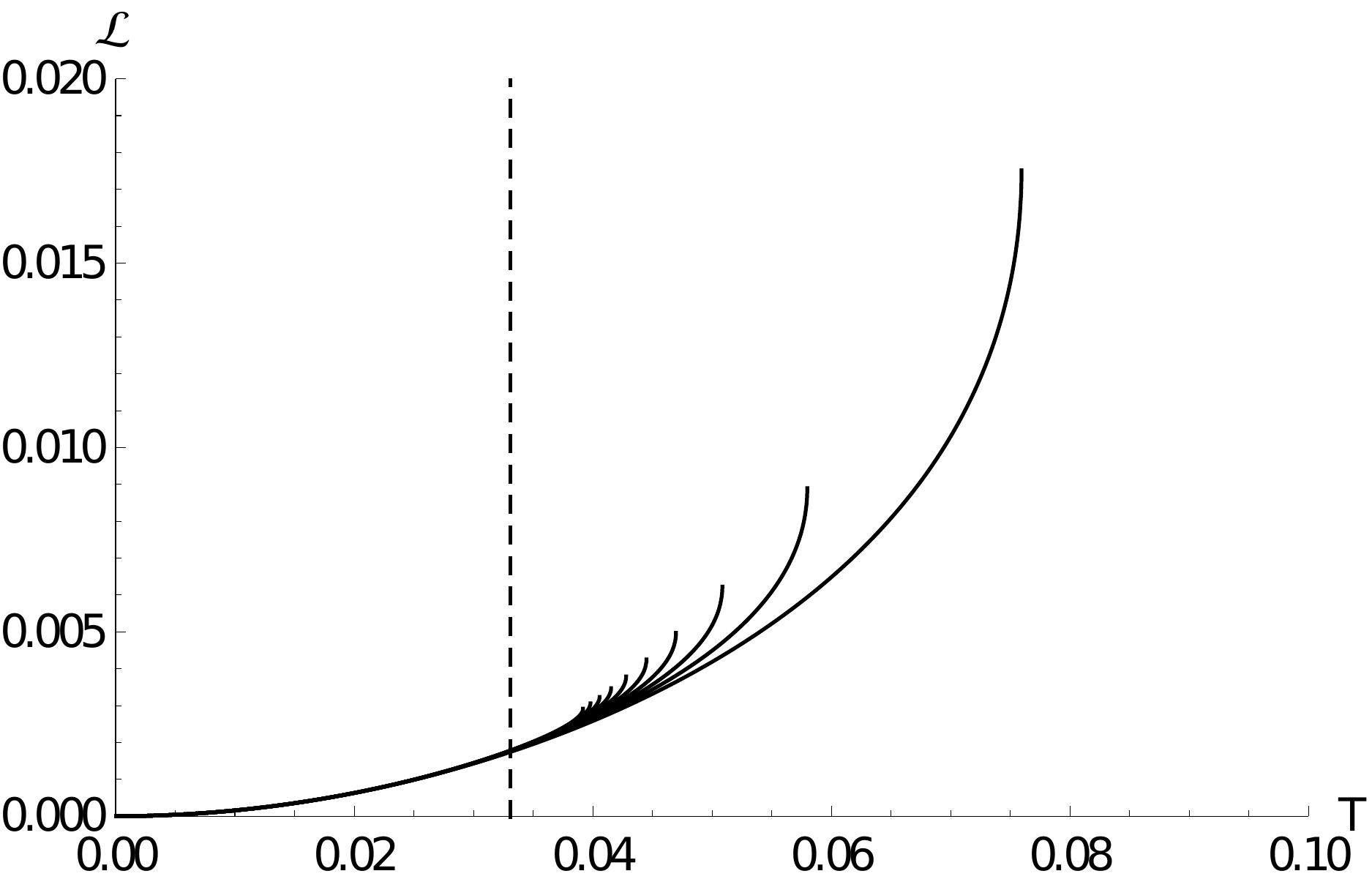}	
	\caption{Spin-4 black hole. BTZ branch for the function $\mc L(T)$  at $\N=4, \dots, 13$, from right to left. The dashed vertical line is placed at the estimated asymptotic value $T_{\rm DS}$.}
	\label{fig:Spin4-Ta}
\end{figure}

\begin{figure}[htbp]
	\centering
	\includegraphics[scale=0.5]{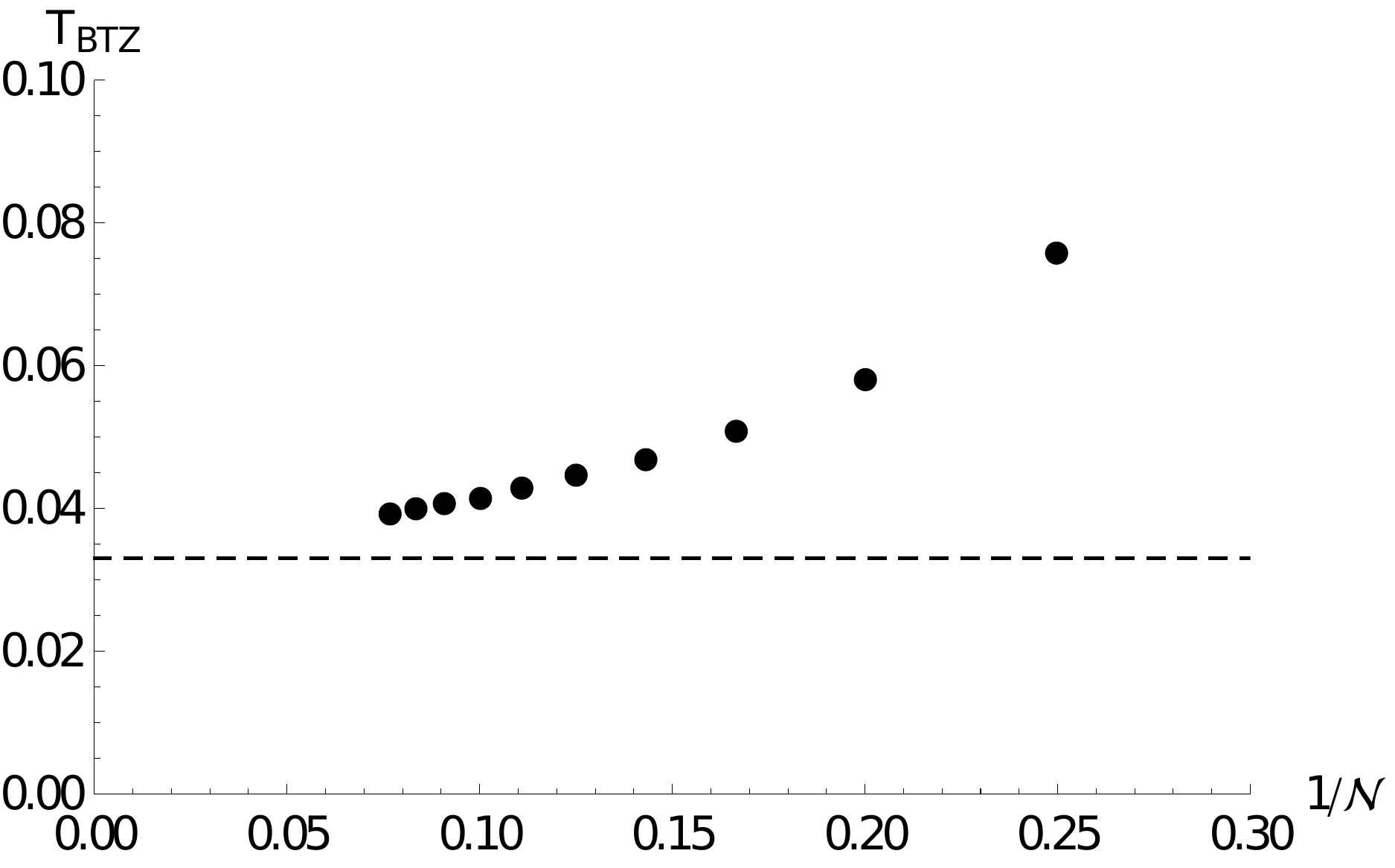}	
	\caption{Spin-4 black hole. Critical BTZ temperature  at $\N=4, \dots, 13$, from right to left. The dashed  line is placed at the estimated asymptotic value $T_{\rm DS}(\N=\infty) $.
}
	\label{fig:Spin4-Tb}
\end{figure}

Again, there is a BTZ branch that stops at a certain $T_{\rm BTZ}(\N)$. This is illustrated
in \figref{Spin4-Ta}, quite similar to \figref{Spin3-Ta}, although with a different position of the vertical dashed line. The explicit values of the critical temperatures are listed in the following table
\be
\begin{array}{|l|l|}
\hline
\N & T_{\rm BTZ}(\N) \\
\hline
4 & 0.0759451227077714684801381449500 \\
5 & 0.0579841118034221349778743376500 \\
6 & 0.0508637951043719636290561268100 \\
7 & 0.0469757284663823535682990022800 \\
8 & 0.0445082433598286822322010852600 \\
\hline
\end{array}
\quad
\begin{array}{|l|l|}
\hline
\N & T_{\rm BTZ}(\N) \\
\hline
9 & 0.0427968534411721973848924582300 \\
10 & 0.0415377037849525320693236372400 \\
11 & 0.0405713402218582096643223297000 \\
12 & 0.0398057427797760377404114516700 \\
13 & 0.0391839637799646669487849202200 \\
\hline
\end{array}
\ee
The convergence of $T_{\rm BTZ}(\N)$ is shown in \figref{Spin4-Tb}, analogous to \figref{Spin3-Tb}. The extrapolation $T_{\rm BTZ}(\infty)$ can be determined by the BST algorithm. We find that the 
best value of the free parameter $\omega$, defined in App.~(\ref{app:BST}), is $\omega^{*}\simeq 0.75$. The BST
estimate reads
\be
\left. T_{\rm BTZ}(\infty)\right|_{\rm Bulirsch-Stoer} = 0.0331(1),
\ee
and is in full agreement with the Domb-Sykes values $T_{\rm DS}$.

\section{Conclusions}

The aim of this paper has been twofold. 
First, we have introduced a novel higher spin black hole solution 
in $\hs$ gravity with a chemical potential coupled to the spin-4 charge. Our approach has been that of working in 
$\mk{sl}(\N)$ gravity and to uplift the results to $\hs$. We derived the expansion of the partition function (and of the various higher spin charges) in powers of the chemical potential. At $\lambda=0,1$, we have been able to test our
results against a free CFT computation. At generic $\lambda$, it would be very interesting to repeat the analysis 
of  \cite{Gaberdiel:2013jca} to see how agreement is found in full details.

The second part of the paper dealt with the thermodynamical properties of the black hole solutions at increasing integer 
$\lambda$. We showed that, both in the standard spin-3 and in the proposed spin-4 solutions, there is a critical 
BTZ temperature $T_{\rm BTZ}(\lambda)$ converging for $\lambda\to\infty$ \cite{Bergshoeff:1989ns} to a finite limit. In the spin-3 case, we 
showed that this asymptotic temperature is associated with the critical point previously found in the partition function 
at $\lambda=\infty$. In the spin-4 case, we showed that the picture is qualitatively similar. By accurate numerical methods,
 we determined the value of $T_{\rm BTZ}(\infty)$ proving that, again, it agrees with the radius of convergence of the perturbative expansion of the partition function.

%

\appendix

\section{Notation for the $\mk{sl}(N)$ generators}
\label{app:notation}

The generators of $\mk{sl}(N)$ in the fundamental representation are $\N\times\N$ matrices $V^{s}_{m}$ 
labeled by a spin and a mode index with $s\ge \N$ and $|m|<s$. The generators of the canonical 
$\mk{sl}(2)$ subalgebra have non-zero matrix elements
\be
\left(V^{2}_{0}\right)_{n,n} = \frac{\N+1}{2}-n, \quad
\left(V^{2}_{1}\right)_{n+1,n}  =  -\sqrt{(\N-n)\,n}, \quad
\left(V^{2}_{-1}\right)_{n,n+1} =  \sqrt{(\N-n)\,n}.
\ee
The other generators are built according to 
\be
V^{s}_{m}=(-1)^{s-1-m}\,\frac{(s+m-1)!}{(2s-2)!}\,\mbox{Ad}_{V^{2}_{-1}}^{s-m-1}\left(V^{2}_{1}\right)^{s-1}.
\ee

\section{List of rational functions $\mc J_{s,n}(\N)$}
\label{app:ratfuncs}

In this section we list the explicit expressions of the rational functions in (\ref{eq:expansions}). For the first two charges $\mc J_{2},\mc J_{4}$ the rational functions are  related, due to the integrability constraint, see (\ref{eq:J2J4constraint}).
For $\mc J_{2}$, we have
\be
\mc J_{2,0}=-\frac{k}{8 \pi }, \qquad \mc J_{2,1}=0,
\ee
and then the generic form of the functions $\mc J_{2,n}(\N)$, with $n=2,3,\dots$, is given by:
\be
\mc J_{2,n}(\N) = \frac{k}{\pi}
\frac{\mathcal{N}^2-9}{\left(\mathcal{N}^2-4\right)^{n-1}} c_{2,n} P_n(\mathcal{N})
\ee
where  the polynomials $P_n(\mathcal{N})$ are 
{\footnotesize
\begin{eqnarray}
P_2&=& \,\mathcal{N}^2-19 , \nonumber\\
P_3&=& 6 \,\mathcal{N}^4-203 \,\mathcal{N}^2+1841 , \nonumber\\
P_4&=& 11 \,\mathcal{N}^6-652 \,\mathcal{N}^4+13338 \,\mathcal{N}^2-92249 , \nonumber\\
P_5&=& 219 \,\mathcal{N}^8-18819 \,\mathcal{N}^6+647154
   \,\mathcal{N}^4-10214359 \,\mathcal{N}^2+60561949 , \nonumber\\
P_6&=& 1536 \,\mathcal{N}^{10}-182045 \,\mathcal{N}^8+9298710
   \,\mathcal{N}^6-248796240 \,\mathcal{N}^4+3383195530
   \,\mathcal{N}^2-18208625939 , \nonumber\\
P_7&=& 80661 \,\mathcal{N}^{12}-12491829 \,\mathcal{N}^{10}+880306840
   \,\mathcal{N}^8-35121988755 \,\mathcal{N}^6+811773363240
   \,\mathcal{N}^4-
\nonumber\\&&
10024855124284 \,\mathcal{N}^2+50497336769391 , \nonumber\\
P_8&=& 25311 \,\mathcal{N}^{14}-4957953 \,\mathcal{N}^{12}+459303411
   \,\mathcal{N}^{10}-25382973765 \,\mathcal{N}^8+876970502835
   \,\mathcal{N}^6-
\nonumber\\&&
18420055913469 \,\mathcal{N}^4+212664969522547
   \,\mathcal{N}^2-1020906501750709 , \nonumber\\
P_9&=& 5157621 \,\mathcal{N}^{16}-1244042667 \,\mathcal{N}^{14}+146286983493
   \,\mathcal{N}^{12}-10663638999384 \,\mathcal{N}^{10}+511505520861370
   \,\mathcal{N}^8-
\nonumber\\&&
16074813998708199 \,\mathcal{N}^6+315502329306026553
   \,\mathcal{N}^4-3465461080026919302
   \,\mathcal{N}^2+
\nonumber\\&&
16038334613159801411 , \nonumber\\
P_{10}&=& 43348788 \,\mathcal{N}^{18}-12613221348
   \,\mathcal{N}^{16}+1833295582573 \,\mathcal{N}^{14}-170175139698053
   \,\mathcal{N}^{12}+
\nonumber\\&&
10791467710583623
   \,\mathcal{N}^{10}-471373487645191887
   \,\mathcal{N}^8+13841886108589726427
   \,\mathcal{N}^6-
\nonumber\\&&
258188309498734124227
   \,\mathcal{N}^4+2728801201784225710397
   \,\mathcal{N}^2-12270580660311002812677 , \nonumber\\
P_{11}&=& 373875271 \,\mathcal{N}^{20}-129036127990
   \,\mathcal{N}^{18}+22700524809270
   \,\mathcal{N}^{16}-2612361851336930
   \,\mathcal{N}^{14}+
\nonumber\\&&
211415628701703860
   \,\mathcal{N}^{12}-12223614161468484033
   \,\mathcal{N}^{10}+499042284927517803085
   \,\mathcal{N}^8-
\nonumber\\&&
13918350328643245849780
   \,\mathcal{N}^6+249470428591687303753995
   \,\mathcal{N}^4-2556920332264806903099715
   \,\mathcal{N}^2+
\nonumber\\&&
11233254880839323385656071 , \nonumber\\
P_{12}&=& 131833590 \,\mathcal{N}^{22}-53221761593
   \,\mathcal{N}^{20}+11141112750572
   \,\mathcal{N}^{18}-1556218041912702
   \,\mathcal{N}^{16}+
\nonumber\\&&
156470911563895957
   \,\mathcal{N}^{14}-11560639992369247612
   \,\mathcal{N}^{12}+625038730778776157706
   \,\mathcal{N}^{10}-
\nonumber\\&& 
24229709614090193714603
   \,\mathcal{N}^8+648768868368037849785923
   \,\mathcal{N}^6-11260481542076455071052893
   \,\mathcal{N}^4+
\nonumber\\&&   
112552589787965953466039948
   \,\mathcal{N}^2-485070302952408138264346901 , \nonumber\\
P_{13}&=& 108550695253 \,\mathcal{N}^{24}-50658748999859
   \,\mathcal{N}^{22}+12441179212060753
   \,\mathcal{N}^{20}-2073371146478408315
   \,\mathcal{N}^{18}+
\nonumber\\&&
253557720211745403685
   \,\mathcal{N}^{16}-23305692981728634874474
   \,\mathcal{N}^{14}+1611172573264592442732782
   \,\mathcal{N}^{12}-
\nonumber\\&& 
82710313469982046488800314
   \,\mathcal{N}^{10}+3076352460580728720146082635
   \,\mathcal{N}^8-
\nonumber\\&&
79680321365763415641963276515
   \,\mathcal{N}^6+1346767858074353279897569700473
   \,\mathcal{N}^4-
\nonumber\\&&
13182874935695491957657605066059
   \,\mathcal{N}^2+55905962753687103934410413898883 \,,
\end{eqnarray}}
and the coefficients $c_{2}$ read
\be
\begin{array}{|c|cccccccccccc|}
\hline
n & 2 & 3 & 4 & 5 & 6 & 7 & 8 & 9 & 10 & 11 & 12 & 13 \\ 
\hline
&&&&&&&&&&&& \\
c_{2,n} & -\frac{3}{2} & 10 & -26 & 192 & -152 & 352 & -\frac{800}{7} & 6400 & 
-\frac{3968}{7} & \frac{8704}{7} &  -\frac{18944}{7} & \frac{1024000}{7} \\
&&&&&&&&&&&& \\
\hline
\end{array}
\ee
For the other charges we have:
\be
\mathcal{J}_{s,\frac{s}{2}-1+k}= \frac{ P_{s,\frac{s}{2}-1+k}}{\left(\mathcal{N}^2-4\right)^{\frac{s}{2}-1+k}},\qquad\qquad
k=0,1,2,\dots\,\, ,
\ee
where the polynomials are given by:

\subsection{$\mathcal{J}_6$ charge}

{\footnotesize
\begin{eqnarray}
&P_{6,2}=&11, \nonumber\\
&P_{6,3}=& -\frac{44}{3} \left(12 \,\mathcal{N}^2-403\right) , \nonumber\\
&P_{6,4}=& 44 \left(78 \,\mathcal{N}^4-4764 \,\mathcal{N}^2+83933\right) , \nonumber\\
&P_{6,5}=& -4400 \left(14 \,\mathcal{N}^6-1363 \,\mathcal{N}^4+49472
   \,\mathcal{N}^2-649951\right) , \nonumber\\
&P_{6,6}=& \frac{176}{3} \left(19389 \,\mathcal{N}^8-2637364
   \,\mathcal{N}^6+152135999 \,\mathcal{N}^4-4252401229
   \,\mathcal{N}^2+47014737269\right) , \nonumber\\
&P_{6,7}=& -3520 \left(6006 \,\mathcal{N}^{10}-1079520 \,\mathcal{N}^8+88179135
   \,\mathcal{N}^6-3956656290 \,\mathcal{N}^4+94344749655
   \,\mathcal{N}^2-930552234394\right) , \nonumber\\
&P_{6,8}=& \frac{10560}{7} \left(264306 \,\mathcal{N}^{12}-60142404
   \,\mathcal{N}^{10}+6519991275 \,\mathcal{N}^8-417136536915
   \,\mathcal{N}^6+16075191384350 \,\mathcal{N}^4-
\right.\nonumber\\&&\left.
344125755485099
   \,\mathcal{N}^2+3128826941308151\right) , \nonumber\\
&P_{6,9}=& -\frac{5632}{3} \left(4054056 \,\mathcal{N}^{14}-1134423548
   \,\mathcal{N}^{12}+156429543836 \,\mathcal{N}^{10}-13349194908765
   \,\mathcal{N}^8+
\right.\nonumber\\&&\left.
737021864713660 \,\mathcal{N}^6-25600471876898424
   \,\mathcal{N}^4+506666627172788252
   \,\mathcal{N}^2-4333421927254544459\right) , \nonumber\\
&P_{6,10}=& \frac{14080}{7} \left(73069689 \,\mathcal{N}^{16}-24584449428
   \,\mathcal{N}^{14}+4183428646012 \,\mathcal{N}^{12}-455943107530981
   \,\mathcal{N}^{10}+
\right.\nonumber\\&&\left.
33704041665411705
   \,\mathcal{N}^8-1682341831113849941
   \,\mathcal{N}^6+54139111900877121927
   \,\mathcal{N}^4-
\right.\nonumber\\&&\left.
1009125696873461387518
   \,\mathcal{N}^2+8228948345347687907799\right) , \nonumber\\
&P_{6,11}=& -\frac{11264}{7} \left(1782317306 \,\mathcal{N}^{18}-708437646651
   \,\mathcal{N}^{16}+145404632545851
   \,\mathcal{N}^{14}-19624311623157486
   \,\mathcal{N}^{12}+
\right.\nonumber\\&&\left.
1858644124109716076
   \,\mathcal{N}^{10}-124530898056055297044
   \,\mathcal{N}^8+5768955965043998717899
   \,\mathcal{N}^6-
\right.\nonumber\\&&\left.
175013595186917334673874
   \,\mathcal{N}^4+3111051322714477655935164
   \,\mathcal{N}^2-24416708826685603901242649\right) , \nonumber\\
&P_{6,12}=& \frac{1024}{21} \left(1158640865382
   \,\mathcal{N}^{20}-536336690469580
   \,\mathcal{N}^{18}+130412248181584215
   \,\mathcal{N}^{16}-
\right.\nonumber\\&&\left.
21294904442849183685
   \,\mathcal{N}^{14}+2504591859141218957495
   \,\mathcal{N}^{12}-215485302317879874185136
   \,\mathcal{N}^{10}+
\right.\nonumber\\&&\left.
13420176964164461096959320
   \,\mathcal{N}^8-586591826385751882753975885
   \,\mathcal{N}^6+
\right.\nonumber\\&&\left.
16975714422410012254404540165
   \,\mathcal{N}^4-290359000474201709856367894530
   \,\mathcal{N}^2+
\right.\nonumber\\&&\left.
2208387891783926147145733884607 
\frac{\phantom{}}{\phantom{}} \right) , \nonumber\\
&P_{6,13}=& -\frac{4096}{7} \left(1918345708356
   \,\mathcal{N}^{22}-1021942827486504
   \,\mathcal{N}^{20}+290129231204176980
   \,\mathcal{N}^{18}-
\right.\nonumber\\&&\left.
56281134810770350635
   \,\mathcal{N}^{16}+8029095210487532534430
   \,\mathcal{N}^{14}-859574130204256381723578
   \,\mathcal{N}^{12}+
\right.\nonumber\\&&\left.
68839878704249084749265982
   \,\mathcal{N}^{10}-4048569504984944304642969120
   \,\mathcal{N}^8+
\right.\nonumber\\&&\left.
168861736037388069692500275990
   \,\mathcal{N}^6-4701446678261012376598318579470
   \,\mathcal{N}^4+
\right.\nonumber\\&&\left.
77888132875458459788739822669366
   \,\mathcal{N}^2-577047049286266867888251654259589\right) 
\end{eqnarray}}

\subsection{$\mathcal{J}_8$ charge}

{\footnotesize
\begin{eqnarray}
&P_{8,3}=& 160 , \nonumber\\
&P_{8,4}=& -120 \left(34 \,\mathcal{N}^2-1721\right) , \nonumber\\
&P_{8,5}=& 160 \left(612 \,\mathcal{N}^4-57206 \,\mathcal{N}^2+1585157\right) , \nonumber\\
&P_{8,6}=& -\frac{800}{3} \left(7956 \,\mathcal{N}^6-1140042
   \,\mathcal{N}^4+63052673 \,\mathcal{N}^2-1301155829\right) , \nonumber\\
&P_{8,7}=& 1280 \left(34986 \,\mathcal{N}^8-6854386 \,\mathcal{N}^6+585653876
   \,\mathcal{N}^4-24876206271 \,\mathcal{N}^2+428626161731\right) , \nonumber\\
&P_{8,8}=& -320 \left(2902716 \,\mathcal{N}^{10}-734612270
   \,\mathcal{N}^8+86721345260 \,\mathcal{N}^6-5751949587190
   \,\mathcal{N}^4+
\right.\nonumber\\&&\left.
207170667267805
   \,\mathcal{N}^2-3155053566516009\right) , \nonumber\\
&P_{8,9}=& \frac{2560}{21} \left(157017168 \,\mathcal{N}^{12}-49358957802
   \,\mathcal{N}^{10}+7568241995020 \,\mathcal{N}^8-698742495681915
   \,\mathcal{N}^6+
\right.\nonumber\\&&\left.
39606752601917970
   \,\mathcal{N}^4-1271531784014296217
   \,\mathcal{N}^2+17688781586885071408\right) , \nonumber\\
&P_{8,10}=& -\frac{2560}{21} \left(3227381388 \,\mathcal{N}^{14}-1226298153654
   \,\mathcal{N}^{12}+234631528568703
   \,\mathcal{N}^{10}-28288801525403720
   \,\mathcal{N}^8+
\right.\nonumber\\&&\left.
2244370698575403930
   \,\mathcal{N}^6-113981294608988239902
   \,\mathcal{N}^4+3358588852763694633371
   \,\mathcal{N}^2-
\right.\nonumber\\&&\left.
43585616773759682215332\frac{\phantom{}}{\phantom{}} \right) , \nonumber\\
&P_{8,11}=& \frac{512000}{7} \left(110576466 \,\mathcal{N}^{16}-49747770072
   \,\mathcal{N}^{14}+11546718838718
   \,\mathcal{N}^{12}-1744763826061109
   \,\mathcal{N}^{10}+
\right.\nonumber\\&&\left.
181544848601967795
   \,\mathcal{N}^8-12954466748658092504
   \,\mathcal{N}^6+605824823897111948798
   \,\mathcal{N}^4-
\right.\nonumber\\&&\left.
16698107749007120988627
   \,\mathcal{N}^2+205063120642839465505911\right) , \nonumber\\
&P_{8,12}=& -\frac{10240}{21} \left(341489933652
   \,\mathcal{N}^{18}-179000572098642
   \,\mathcal{N}^{16}+49346093247710917
   \,\mathcal{N}^{14}-
\right.\nonumber\\&&\left.
9078583420393849712
   \,\mathcal{N}^{12}+1188124194558471978367
   \,\mathcal{N}^{10}-111536090354401467491548
   \,\mathcal{N}^8+
\right.\nonumber\\&&\left.
7347700985298618829454333
   \,\mathcal{N}^6-322071096020723601086168158
   \,\mathcal{N}^4+
\right.\nonumber\\&&\left.
8413233300038269921929300963
   \,\mathcal{N}^2-98779213733726462824355763708\right) 
\end{eqnarray}}

\subsection{$\mathcal{J}_{10}$ charge}

{\footnotesize
\begin{eqnarray}
&P_{10,4}=& 2660 , \nonumber\\
&P_{10,5}=& -\frac{304}{3} \left(924 \,\mathcal{N}^2-64481\right) , \nonumber\\
&P_{10,6}=& \frac{1520}{9} \left(15939 \,\mathcal{N}^4-2076957
   \,\mathcal{N}^2+81193304\right) , \nonumber\\
&P_{10,7}=& -\frac{30400}{3} \left(6699 \,\mathcal{N}^6-1313118
   \,\mathcal{N}^4+101115967 \,\mathcal{N}^2-2951553716\right) , \nonumber\\
&P_{10,8}=& \frac{6080}{21} \left(5567331 \,\mathcal{N}^8-1472721481
   \,\mathcal{N}^6+172424322021 \,\mathcal{N}^4-10178523096191
   \,\mathcal{N}^2+247319714365476\right) , \nonumber\\
&P_{10,9}=& -\frac{24320}{63} \left(95496786 \,\mathcal{N}^{10}-32195517795
   \,\mathcal{N}^8+5133311615460 \,\mathcal{N}^6-465576474795240
   \,\mathcal{N}^4+
\right.\nonumber\\&&\left.
23220347933875530
   \,\mathcal{N}^2-496201639994999189\right) , \nonumber\\
&P_{10,10}=& \frac{24320}{63} \left(2135187054 \,\mathcal{N}^{12}-883423972431
   \,\mathcal{N}^{10}+180538974433860 \,\mathcal{N}^8-22461662733198920
   \,\mathcal{N}^6+
\right.\nonumber\\&&\left.
1734758130575491210
   \,\mathcal{N}^4-76771604427700863401
   \,\mathcal{N}^2+1490341998314655353124\right) , \nonumber\\
&P_{10,11}=& -\frac{97280}{21} \left(3920605689 \,\mathcal{N}^{14}-1938905021662
   \,\mathcal{N}^{12}+488479935450759
   \,\mathcal{N}^{10}-78315272143852160
   \,\mathcal{N}^8+
\right.\nonumber\\&&\left.
8343508297358717665
   \,\mathcal{N}^6-574944227639131582056
   \,\mathcal{N}^4+23242634786878229150663
   \,\mathcal{N}^2-
\right.\nonumber\\&&\left.
418615852040056581869346\frac{\phantom{}}{\phantom{}}\right) 
\end{eqnarray}
}

\subsection{$\mathcal{J}_{12}$ charge}

{\footnotesize
\begin{eqnarray}
&P_{12,5}=& 47840 , \nonumber\\
&P_{12,6}=& -\frac{18400}{3} \left(351 \,\mathcal{N}^2-31894\right) , \nonumber\\
&P_{12,7}=& 14720 \left(4914 \,\mathcal{N}^4-840757
   \,\mathcal{N}^2+43400654\right) , \nonumber\\
&P_{12,8}=& -\frac{36800}{21} \left(1184274 \,\mathcal{N}^6-301767018
   \,\mathcal{N}^4+30503260417 \,\mathcal{N}^2-1179416152291\right) , \nonumber\\
&P_{12,9}=& \frac{29440}{189} \left(351242892 \,\mathcal{N}^8-119791374192
   \,\mathcal{N}^6+18233821986597 \,\mathcal{N}^4-
\right.\nonumber\\&&\left.
1411369274423162
   \,\mathcal{N}^2+45381731381100032\right) , \nonumber\\
&P_{12,10}=& -\frac{29440}{21} \left(974932686 \,\mathcal{N}^{10}-420139561170
   \,\mathcal{N}^8+86292851439585 \,\mathcal{N}^6-10157226055501115
   \,\mathcal{N}^4+
\right.\nonumber\\&&\left.
662741868540001530
   \,\mathcal{N}^2-18690347154396001414\right) 
\end{eqnarray}}

\subsection{$\mathcal{J}_{14}$ charge}

{\footnotesize
\begin{eqnarray}
&P_{14,6}=&906192 , \nonumber\\
&P_{14,7}=& -4608 \left(10788 \,\mathcal{N}^2-1226597\right) , \nonumber\\
&P_{14,8}=& 10752 \left(178002 \,\mathcal{N}^4-38356776
  \,\mathcal{N}^2+2500615697\right) , \nonumber\\
&P_{14,9}=& -\frac{8960}{3} \left(20583504 \,\mathcal{N}^6-6569096328
   \,\mathcal{N}^4+836392234032 \,\mathcal{N}^2-40958747755511\right)
\end{eqnarray}}

\subsection{$\mathcal{J}_{16}$ charge}

{\footnotesize
\begin{eqnarray}
&P_{16,7}=&17808384 , \nonumber\\
&P_{16,8}=& -\frac{3968}{3} \left(871794 \,\mathcal{N}^2-120447311\right) 
\end{eqnarray}}

\section{Properties of the zeroes of $\mc J_{s,n}(\N)$}
\label{app:zeroes}

In \cite{Beccaria:2013dua}, it was observed that the polynomials appearing in the expansion of the charges had a certain number of real roots moving toward integer numbers as the degree of the polynomial increased. Here, we can repeat the same numerical analysis for the polynomials $P_m$ in the expansion of $\mathcal{L}$. Again, we find some roots approaching integer numbers, as one can clearly see in the following table where we report for each polynomial the nearest real root (when present) to $4,5,\dots$.

{\footnotesize
\be
\begin{array}{c|c|c|c|c|c}
P_m \, \,&\,  \textrm{nearest root to 4}  \,&\, \textrm{nearest root to 5}  \,&\, \textrm{nearest root to 6}  \,&\, \textrm{nearest root to 7}  \,&\, \textrm{nearest root to 8} \\
\hline
P_{2}\, \,&\, 4.35889894354067355 \,&\,                                           \,&\,                                           \,&\,  \,&\, \\
P_{3}\, \,&\,                                       \,&\,                                           \,&\,                                           \,&\,  \,&\, \\
P_{4}\, \,&\,4.10423499541088212  \,&\,                                           \,&\,                                           \,&\,  \,&\, \\
P_{5}\, \,&\,4.02583845153139326 \,&\, 4.69606738643800 \,&\,                                           \,&\,  \,&\, \\
P_{6}\, \,&\,4.00255705894313841 \,&\,                                           \,&\,                                           \,&\,  \,&\, \\
P_{7}\, \,&\,4.00026667286335657 \,&\, 4.95603816837181 \,&\,                                           \,&\,  \,&\, \\
P_{8}\, \,&\,4.00001737732443553 \,&\, 5.00903114953433 \,&\, 5.59430121473374 \,&\,  \,&\, \\
P_{9}\, \,&\,4.00000103167653541 \,&\, 4.99890565769590 \,&\,                                           \,&\,  \,&\, \\
P_{10}\, \,&\,4.00000004575481496 \,&\, 5.00010499793752 \,&\, 5.91805274953158 \,&\,  \,&\,  \\
P_{11}\, \,&\,4.00000000179973362 \,&\, 4.99999128613680 \,&\, 6.02757081000280 \,&\, 6.43588276581036 \,&\,  \\
P_{12}\, \,&\,4.00000000005775033 \,&\, 5.00000058476395 \,&\, 5.99679058314277 \,&\,                                           \,&\,  \\
P_{13}\, \,&\,4.00000000000163800 \,&\, 4.99999996556779 \,&\, 6.00040130652251 \,&\, 6.86165207315471 \,&\,         \\
P_{14}\, \,&\,4.00000000000003972 \,&\, 5.00000000172409 \,&\, 5.99995851361281 \,&\, 7.10362284961468 \,&\, 7.22004539884681  \\
P_{15}\, \,&\,4.00000000000000085 \,&\, 4.99999999992356 \,&\, 6.00000376500671 \,&\, 6.99199759082295 \,&\,          \\
P_{16}\, \,&\,4.00000000000000001 \,&\, 5.00000000000296 \,&\, 5.99999970088932 \,&\, 7.00115961192149 \,&\, 7.78990734987434 \\
\hline
\end{array}
\ee}
The interpretation of this property is the same as in \cite{Beccaria:2013dua}. In particular, 
this property is related to the  the truncation of
the underlying $\mc W_{\infty}$ algebra that happens precisely at the integer values of $\lambda$.

\section{High order expansion of $\log Z_{\N=\infty}$}
\label{app:expansion}

The expansions coefficients $b_{n}\equiv b_{n}(\N=\infty)$ appearing in (\ref{eq:Zexp}) can be computed by 
the formula (\ref{eq:Ninftyresult}). We found
\ba
b_{0} &=& 1, \nonumber\\
b_{1} &=& 0, \nonumber\\
b_{2} &=& 12/7, \nonumber\\
b_{3} &=& -8, \nonumber\\
b_{4} &=& 96, \nonumber\\
b_{5} &=& -1056, \nonumber\\
b_{6} &=& 14016, \nonumber\\
b_{7} &=& -196608, \nonumber\\
b_{8} &=& 2949888, \nonumber\\
b_{9} &=& -323980800/7, \nonumber\\
b_{10} &=& 754486272, \nonumber\\
b_{11} &=& -12682616832, \nonumber\\
b_{12} &=& 218770444288, \nonumber\\
b_{13} &=& -3857074176000, \nonumber\\
b_{14} &=& 69291997052928, \nonumber\\
b_{15} &=& -1265276167618560, \nonumber\\
b_{16} &=& 164054123598249984/7, \nonumber\\
b_{17} &=& -439613746473861120, \nonumber\\
b_{18} &=& 8339276221242408960, \nonumber\\
b_{19} &=& -159790677648856842240, \nonumber\\
b_{20} &=& 3089636074275669540864, \nonumber\\
b_{21} &=& -60231583326427391459328, \nonumber\\
b_{22} &=& 1182989504014215404322816, \nonumber\\
b_{23} &=& -163755327799352588073172992/7, \nonumber\\
b_{24} &=& 465509751499617144548622336, \nonumber\\
b_{25} &=& -9316638506633894891278565376, \nonumber\\
b_{26} &=& 187455304711114086670223278080, \nonumber\\
b_{27} &=& -3790306305519518769991737409536, \nonumber\\
b_{28} &=& 76990497341210622205257452617728, \nonumber\\
b_{29} &=& -1570543496736509871027382476865536, \nonumber\\
b_{30} &=& 225158429924385360071748400025960448/7, \nonumber\\
b_{31} &=& -661221361804071797940248549932400640, \nonumber\\
b_{32} &=& 13640163143067197235022990584311709696, \nonumber\\
b_{33} &=& -282303859532708338543828142820239081472.
\ea

\section{The Bulirsch-Stoer extrapolation algorithm}
\label{app:BST}

An interesting problem in numerical mathematics is that of estimating the asymptotic value 
$f_\infty$ of a sequence $\{f_n\}$, given a finite number of terms. The problem is easy when the sequence is 
linearly convergent, {\em i.e.} satisfies the condition
\be
|\rho| < 1,\qquad\ {\rm with}\ \ \ \rho = \lim_{n\to\infty}\frac{f_{n+1}-f_\infty}{f_n-f_\infty}  .
\ee 
In this case rigorous theorems provide the existence of algorithms that accelerate 
the asymptotic convergence of any such sequence. In the logarithmically convergent case, $\rho=1$,
there is not any method that can guarantee the acceleration of  a generic sequence. However, efficient acceleration algorithms exist if if one selects a restricted class of logarithmically convergent sequences. A widely considered 
case is that of sequences $f_n\sim f_\infty + a_1 n^{-\omega_1} + a_2 n^{-\omega_2} + \dots$
where $0<\omega_1<\omega_2<\dots$ are positive exponents. They have  $\rho=1$, but can be treated with the 
so-called BST algorithm~\cite{BST}. It is convenient to introduce a generic small parameter $h$ and consider 
a function $f(h)$ with the following asymptotic expansion as $h\to 0$
\be
f(h) = f(0) + a_1 h^{\omega_1} + a_2 h^{\omega_2} + \cdots,\qquad 0 < \omega_1 < \omega_2 < \cdots.
\ee
Now, suppose that the following set of pairs is available
\be
\{(h_n, f(h_n))\}_{0\le n < N},\qquad 0 < h_N < h_{N-1} < \cdots < h_0 .
\ee
The BST algorithm estimates $f(0)$ by constructing a  improved sequences whose convergence is 
accelerated with respect to the initial sequence. Technically, one builds a grid 
 $\{f_{n,m}\}_{0\le n < N, -1\le m < N}$ where the initial values are 
\be
f_{n,-1}=0,\qquad f_{n,0} = f(h_n),
\ee
and the next ones are computed by iterating with respect to $m$ the relation
\be
f_{n,m} = f_{n+1,m-1}+(f_{n+1,m-1}-f_{n,m-1})\left[\left(\frac{h_n}{h_{n+m}}\right)^\omega
\left(1-\frac{f_{n+1,m-1}-f_{n,m-1}}{f_{n+1,m-1}-f_{n+1,m-2}}\right)-1\right]^{-1} ,
\ee
where $\omega$ is a positive real number which is a free parameter of the algorithm.

After that the BST grid has been computed for a certain $\omega$, the best choice $\omega=\omega^*$ is the one that makes the last generated sequences ({\em i.e.} $f_{n,m}$ with $m=N-1$, $N-2$, $\dots$) as flat as possible.
A measure of the ``non-flatness'' of the last sequences is typically built choosing a small integer $K$ and considering
the border sum of differences
\be
\label{delta}
\delta_K(\omega) = 
\sum_{j=0}^{K-1} |f_{j,N-1-j}(\omega)-f_{j + 1, N-2-j}(\omega)| + \sum_{j=1}^K |f_{j, N-1-j}(\omega)-f_{j - 1, N-1-j}(\omega)| ,
\ee
Then, $\omega^*$ as the value that minimize $\delta_K(\omega)$, and the predicted limiting value is 
$f_{0,N-1}(\omega^*)$.

\providecommand{\href}[2]{#2}\begingroup\raggedright\endgroup

\end{document}